\documentclass[a4paper,11pt]{article}

\usepackage{jheppub}
\usepackage{graphicx}
\usepackage{amsmath}
\usepackage{slashed}
\usepackage{braket}
\usepackage[export]{adjustbox}
\usepackage{enumitem}
\usepackage{placeins}
\usepackage[normalem]{ulem}

\newcommand{\refcite}[1]{ref.~\cite{#1}}
\newcommand{\refscite}[1]{refs.~\cite{#1}}
\newcommand{\Eq}[1]{Eq.~\eqref{eq:#1}}
\newcommand{\eq}[1]{eq.~\eqref{eq:#1}}
\newcommand{\eqs}[2]{eqs.~\eqref{eq:#1} and \eqref{eq:#2}}
\renewcommand{\sec}[1]{section~\ref{sec:#1}}

\newcommand{\app}[1]{appendix~\ref{app:#1}}
\newcommand{\fig}[1]{figure~\ref{fig:#1}}

\newcommand{\qpsq}{q_T^2}
\newcommand{\df}{\mathrm{d}}
\newcommand{\img}{\mathrm{i}}
\newcommand{\eps}{\epsilon}

\newcommand{\bn}{{\bar n}}

\newcommand{\cA}{\mathcal{A}}
\newcommand{\cB}{\mathcal{B}}
\newcommand{\cC}{\mathcal{C}}
\newcommand{\cF}{\mathcal{F}}
\newcommand{\cI}{\mathcal{I}}

\newcommand{\cL}{\mathcal{L}}
\newcommand{\cK}{\mathcal{K}}
\newcommand{\cM}{\mathcal{M}}
\newcommand{\cN}{\mathcal{N}}
\newcommand{\cO}{\mathcal{O}}

\newcommand{\qt}{{\vec q}_T}
\newcommand{\bt}{{\vec b}_T}

\newcommand{\pt}{{\vec p}_T}

\newcommand{\myzeta}{\xi}
\newcommand{\wa}{{w_1}}
\newcommand{\wb}{{w_2}}

\newcommand{\bnlim}{\lim\limits_{\mathrm{strict}~p_2-\mathrm{coll.}}}
\newcommand{\as}{\alpha_s}
\newcommand{\aem}{\alpha_{\rm em}}

\newcommand{\nn}{\nonumber}

\newcommand{\lqcd}{\Lambda_\mathrm{QCD}}

\newcommand{\blue}[1]{{\color{blue}#1}}
\newcommand{\TMD}{\mathrm{TMD}}

\def\beq{\begin{equation}}
\def\eeq{\end{equation}}
\def\bea{\begin{eqnarray}}
\def\eea{\end{eqnarray}}

\title{\boldmath TMD Fragmentation Functions at N$^3$LO}

\author[a,b]{Markus A.~Ebert,}
\emailAdd{ebert@mpp.mpg.de}

\author[c]{Bernhard Mistlberger,}
\emailAdd{bernhard.mistlberger@gmail.com}

\author[b,c]{and Gherardo Vita}
\emailAdd{gvita@stanford.edu}

\affiliation[a]{Max-Planck-Institut f\"ur Physik, F\"ohringer Ring 6, 80805 M\"unchen, Germany}
\affiliation[b]{Center for Theoretical Physics, Massachusetts Institute of Technology, Cambridge, Massachusetts 02139, USA}
\affiliation[c]{SLAC National Accelerator Laboratory, Stanford University, Stanford, CA 94039, USA}

\abstract{
We compute the unpolarized quark and gluon transverse-momentum dependent fragmentation functions (TMDFFs) at next-to-next-to-next-to-leading order (N$^3$LO) in perturbative QCD.
The calculation is based on a relation between the TMDFF and the limit of the semi-inclusive deep inelastic scattering cross section where all final-state radiation becomes collinear to the detected hadron.
The required cross section is obtained by analytically continuing our recent computation of the Drell-Yan and Higgs boson production cross section at N$^3$LO expanded around the limit of all final-state radiation becoming collinear to one of the initial states.
Our results agree with a recent independent calculation by Luo et al.
}

\preprint{\vbox{%
\hbox{MIT--CTP 5261}
\hbox{SLAC--PUB--17577}
\hbox{MPP--2020--220}
}}

\begin{document}

\maketitle
\newpage

\section{Introduction}
\label{sec:intro}

Highly energetic scattering processes allow us to test our understanding of fundamental interactions with incredible precision.
It is the asymptotic freedom of strong interactions of QCD that allows us to contrast our first principle understanding of the interactions with experimental data.
The interacting elementary particles of QCD - quarks and gluons - are however concealed in our observations as they form hadronic bound states as the strong interactions confine at long distances.
The gateway that bridges the world of partonic interactions, where observables are calculable in perturbative QCD, to the observation in real-life detectors is provided by factorisation theorems.
Factorisation theorems split the long range, confining part of a scattering process from the short range collision process of quarks and gluons. 
The long range part of this factorisation theorems is typically expressed in terms of parton distribution functions (PDFs) and fragmentation functions (FFs). 
PDFs and FFs are independent of the particularities of the scattering process and are universal, such that they can be measured and used in many different experiments and observables.

Longitudinal FFs are the simplest example of FFs, as they only describe the probability of a quark or a gluon to convert to a hadron that carries a given momentum fraction of the fragmenting parton~\cite{Georgi:1977mg,Ellis:1978ty,Collins:1981uw,Collins:1989gx}.
This notion is expanded by transverse-momentum dependent FFs (TMDFFs)~\cite{Collins:1981uk, Collins:1981va,Collins:1992kk,Mulders:1995dh,Boer:1997nt,Boer:1997qn,Ji:2004wu,Collins:2011zzd},
which encode the probability of a hadron to arise from a fragmenting parton with a certain fraction of the partons longitudinal momentum and a small transverse momentum relative to the parton.

TMDFFs are important ingredients for describing high-energy scattering processes involving hadronic final states at low transverse momentum,
for example hadron production at $e^+e^-$ colliders or semi-inclusive deep-inelastic scattering (SIDIS)~\cite{Ashman:1991cj,Derrick:1995xg,Adloff:1996dy,Aaron:2008ad,Airapetian:2012ki,Adolph:2013stb,Aghasyan:2017ctw},
which will also play an important role at the upcoming Electron-Ion Collider (EIC)~\cite{Accardi:2012qut, Aschenauer:2019kzf}.
This data has been used to extract both unpolarized TMDFFs~\cite{Su:2014wpa, Bacchetta:2017gcc, Scimemi:2019cmh} and the so-called Sivers function~\cite{Efremov:2004tp, Vogelsang:2005cs, Anselmino:2005ea, Anselmino:2008sga, Aybat:2011ta, Gamberg:2013kla, Sun:2013dya, Echevarria:2014xaa, Anselmino:2016uie, Bacchetta:2020gko, Cammarota:2020qcw, Echevarria:2020hpy}.
TMDFFs are also closely related to TMD jet functions arising in processes involving final-state jets at low transverse momentum~\cite{Neill:2016vbi,Gutierrez-Reyes:2018qez,Gutierrez-Reyes:2019vbx,Gutierrez-Reyes:2019msa}, and to the jet functions encountered in energy correlation functions in electron-positron annihilation \cite{Moult:2018jzp,Ebert:2020sfi} and transverse-momentum dependent event shapes involving jets~\cite{Gao:2019ojf,Li:2020bub,Ali:2020ksn}.
For a review on TMDFFs see for example \refcite{Metz:2016swz} and references therein.

TMDFFs are intrinsically nonperturbative objects, as they relate the dynamics of partons and hadrons,
and as such have been extracted from various experiments~\cite{Sun:2013hua,Echevarria:2014xaa,Metz:2016swz,Bacchetta:2017gcc,Bertone:2017tyb,Bertone:2018ecm,Bertone:2019nxa,Callos:2020qtu,Lin:2017stx}.
However, for transverse momenta $q_T$ that are much larger than the confinement scale $\lqcd$, an operator product expansion in $\lqcd/q_T$ allows one to express each TMDFF in terms of a standard longitudinal FF and a $q_T$-dependent matching kernel.
The matching kernels are calculable order by order in perturbation theory and are currently known at next-to-next-to-leading order (NNLO)~\cite{Luo:2019bmw,Luo:2019hmp,Echevarria:2016scs} in perturbative QCD.
In the regime of perturbative $q_T$, they can be used for example in extractions of longitudinal FFs from differential measurements of suitable observables, see for example refs.~\cite{Jain:2011iu,Seidl:2019jei,Makris:2020ltr,Boglione:2020auc}.

In this article we present the calculation of the matching kernels for all unpolarized quark and gluon TMDFFs at N$^3$LO.
TMD parton distribution functions (TMDPDFs), the initial-state counterparts of TMDFFs, are already known at this order~\cite{Ebert:2020yqt,Luo:2019szz}.
Our calculation relies on a recently developed method to expand hadron collider cross sections around the limit where final state QCD radiation is collinear to an incoming parton~\cite{Ebert:2020lxs}.
We demonstrate explicitly how partonic cross sections for the production of electro-weak gauge bosons can be related to DIS cross sections via analytic continuation.
We apply this analytic continuation to the collinear limit of the partonic cross section of gluon-fusion Higgs boson production and Drell-Yan production to obtain their DIS counter part.
The collinear limit of these production cross sections was recently computed by us for the calculation of the N$^3$LO TMDPDFs~\cite{Ebert:2020yqt} and $N$-jettiness beam function~\cite{Ebert:2020unb}.
We then establish an analytic relation among the TMDFF matching kernels and the newly obtained collinear limit of the DIS cross sections.
The combination of this collinear limit with the TMD soft function yields the scheme-independent TMDFF at N$^3$LO.
With this we are finally able to extract the desired perturbative TMDFF matching kernels.

The paper is structured as follow. 
In \sec{setup}, we setup the kinematics for SIDIS retaining full information on the momentum of the final state hadron.
In \sec{crossing}, we show how to use crossing symmetry and analytic continuation to obtain results for fully differential partonic cross sections in SIDIS from analogous cross sections in proton-proton collision.
In \sec{collinear_limit}, we study the behaviour of the partonic cross section when taking the radiation to be collinear either to the struck proton or to the final state hadron.
In \sec{TMDFFExt}, we make use of the framework developed in the previous sections to extract the TMDFFs at N$^3$LO by imposing a transverse-momentum measurement to the leading collinear expansion of the cross section.
We conclude in \sec{conclusion}.

\section{Setup}
\label{sec:setup}

In this section we introduce our notation for the description of semi-inclusive deep inelastic scattering (SIDIS),
reviewing both the scattering process and providing the definitions of all required kinematic variables and the associated final-state phase space.
Finally, we define the transverse momentum observables of interest in this article.

\subsection{Semi Inclusive Deep Inelastic Scattering}
\label{sec:SIDIS}

\begin{figure*}
\centering
\includegraphics[width=0.6\textwidth]{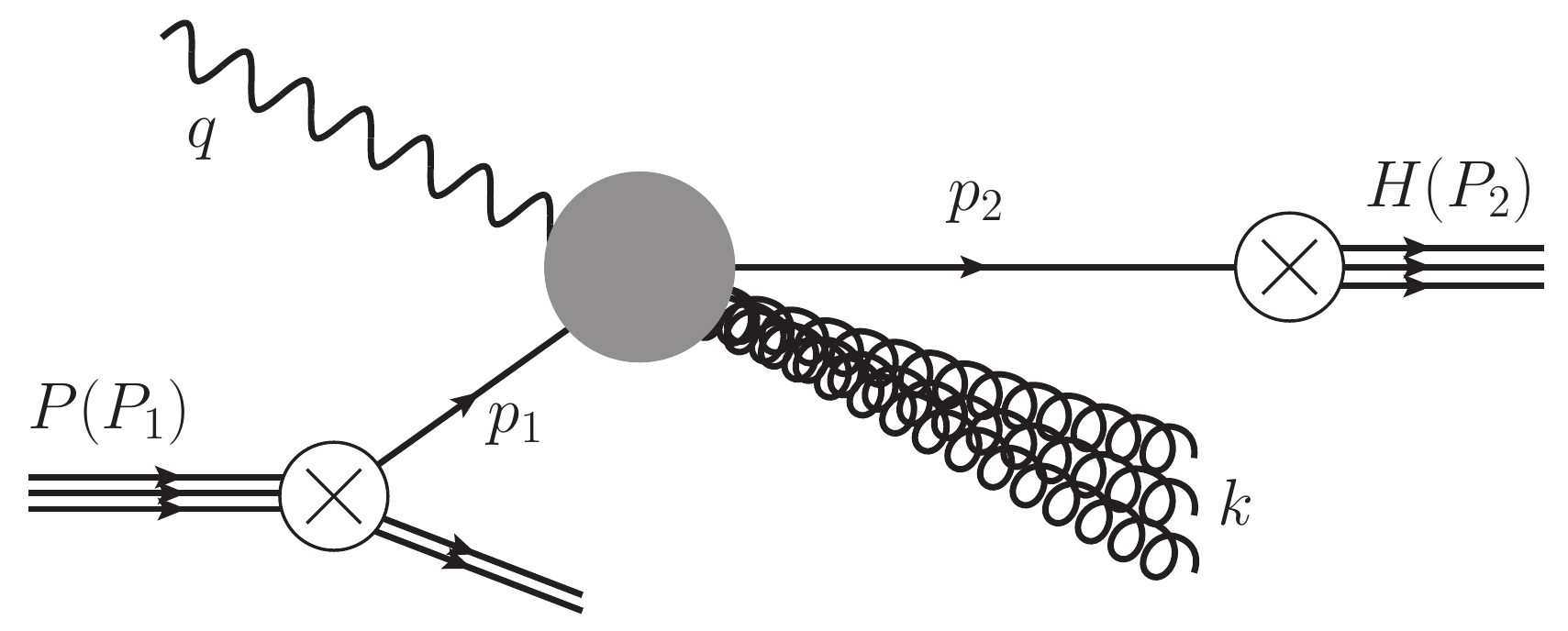}
\caption{Schematic picture of the DIS process in \eq{DIS}, producing a final-state hadron $H$ in the scattering of an electroweak boson, here a photon, off the incoming proton.}
\label{fig:DIS}
\end{figure*}

We study cross sections for the production of a hadron $H$ in DIS alongside additional radiation, which we indicate as a multiparticle state $X$. 
In particular, we focus on the hadronic part of the DIS cross section that is initiated by the scattering of a proton with momentum $P_1$  and an electro-weak boson $h$ with the space-like momentum $q$,
\beq \label{eq:DIS}
 P(P_1) \, + \,h(q) \rightarrow H(-P_2) \,+\,X(-k)
\,.\eeq
Here, we take all momenta to be incoming.
This process is schematically depicted in \fig{DIS} for the example of a virtual photon as the electro-weak gauge boson.
In this article, we will consider DIS with either a virtual photon or a Higgs boson as the electro-weak gauge boson.

We are interested in SIDIS, where we measure an observable $\cO$ that depends on the final-state hadronic momenta.
For perturbative $\cO \gg \lqcd$, the cross section differential in $\cO$ can be factorized as
\begin{align} \label{eq:hadr}
 \frac{\df\sigma_{P+h \rightarrow H+X}}{\df {x_F}\df\cO} &
 = \hat\sigma_0 \sum_{i,j} f_i({x_B})\otimes_{x_B} \frac{\df\hat\eta_{i+h\rightarrow j+X}({x_B},{x_F},\cO)}{\df x_F \df\cO}\otimes_{x_F} d_{H/j}({x_F})
\,.\end{align}
Here, the overall normalization $\hat\sigma_0$ is the Born cross section and the sum runs over parton flavors $i,j$.
In \eq{hadr}, $\hat\eta_{i+h\rightarrow j+X}$ is a perturbatively calculable partonic coefficient function encoding the underlying partonic process $i+h\rightarrow j+X$,
which is convolved with the nonperturbative parton distribution function (PDF) $f_i$ and fragmentation function (FF) $d_{H/j}$.
The PDF $f_i(x)$ encodes the probability to extract the parton with flavor $i$ and momentum fraction $x$ from the proton,
while the FF $d_{H/j}({y})$ describes the fragmentation of a parton of flavor $j$ into a hadron of type $H$ which carries the momentum fraction $y$ of the parent parton.
We define the hadronic invariants
\bea \label{eq:xB_xi}
 x_B = -\frac{q^2}{2P_1 \cdot q} \,,\qquad  {x_F} = -\frac{2P_2\cdot q}{q^2}
\,.\eea
In analogy, we introduce the partonic variables $z$ and $\myzeta$,
\bea
\label{eq:zetadef}
z=-\frac{q^2}{2p_1 \cdot q},\hspace{1cm}\myzeta=-\frac{2p_2 \cdot q}{q^2}.
\eea
The convolution integrals abbreviated by $\otimes_{x_B}$ and $\otimes_{x_F}$ in \eq{hadr} can now be written explicitly as
\begin{align} \label{eq:hadr2}
 \frac{\df\sigma_{P+h \rightarrow H+X}}{\df {x_F}\df\cO} &
 = \hat\sigma_0 \sum_{i,j} \int_{x_B}^1\frac{dz}{z} \int_{x_F}^1\frac{d\myzeta}{\myzeta} f_i\Bigl(\frac{x_B}{z}\Bigr)
   \frac{\df\hat\eta_{i+h\rightarrow j+X}(z,\myzeta,\cO)}{\df \myzeta \df\cO} d_{H/j}\Bigl(\frac{x_F}{\myzeta}\Bigr)
\,,\end{align}
where the partonic coefficient function is given by
\begin{align} \label{eq:partdef}
 \frac{\df\hat\eta_{i+h\rightarrow j+X}}{\df\myzeta\df\cO} &
 = \frac{1}{\hat\sigma_0} \frac{\cN_i}{2|q^2|} \sum\limits_{m=0}^\infty
   \int\!\df\Phi_{1+m} \, \delta\Bigl(\myzeta+\frac{2p_2 \cdot q}{q^2}\Bigr)
   \delta\bigl[\cO - \hat\cO\bigl(p_2,{x_B},z,{x_F},\myzeta\bigr)\bigr] \, |\cM_{i+h\rightarrow j+m}|^2
\,.\end{align}
Here we introduced the normalization factor $\cN_i$ related to the helicity and color average of the incoming particle,
which for an incoming quark or gluon takes the value
\beq
\cN_g=\frac{1}{2(1-\epsilon)(n_c^2-1)}\,,\qquad
\cN_q=\frac{1}{2n_c}\,.
\eeq
In \eq{partdef}, the sum runs over the number $m$ of additional partons in the final state besides the parton of flavor $j$ that fragments into the hadron $H$,
and $\Phi_{1+m}$ is the associated $m+1$-parton phase space.
The $\delta$ functions implement the measurements of $\myzeta$ and $\cO$, and the squared matrix element $|\cM_{i+h\rightarrow j+m}|^2$
corresponds to the partonic process of producing the $m+1$ partons in the collision of a parton of flavor $i$ with the hard probe $h$.

\subsection{Kinematics and Final State Phase Space}
\label{sec:vardef}
We are interested in observables differential in the four momentum $P_2$ of the final state hadron $H$,
while we are inclusive over all additional final-state radiation.
A convenient set of variables to describe the kinematics of the corresponding partonic process is given by
\begin{align} \label{eq:supervars}
 s=(p_1+p_2)^2
\,,\quad
 w_1=-\frac{2p_1 \cdot k}{2p_1 \cdot p_2}
\,,\quad
 w_2=-\frac{2p_2 \cdot k}{2p_1 \cdot p_2}
\,,\quad
 x=\frac{(2p_1 \cdot p_2) k^2}{(2p_1 \cdot k)(2p_2 \cdot k)}
\,.\end{align}
Here, $k$ is the sum of all $m$ final state momenta of the particles produced in addition to the parton with momentum $p_2$,
\beq
k=\sum_{i=3}^{m+2}p_i \,.
\eeq
The differential $m+1$-parton phase space is given by
\beq
\df\Phi_{1+m}=(2\pi)^d \delta^d\left(p_1+q+\sum_{i=2}^{m+2}p_i\right) \prod_{i=2}^{m+2} \frac{\df^dp_i}{(2\pi)^d} (2\pi)\delta_+(p_i^2)\,.
\eeq
It can be parameterized using the variables in \eq{supervars} as
\begin{align} \label{eq:phi_1m}
 \frac{\df \Phi_{1+m}}{ \df w_1\df w_2 \df x} &
 = \frac{\Omega_{2-2\epsilon}}{4(2\pi)^{3-2\epsilon}} (q^2 w_1 w_2)^{1-\epsilon} (1-w_1)^{-3+2\epsilon} (1-x)^{-\epsilon} (1-w_1-w_2+w_1 w_2 x)^{-1+\epsilon}
\nn\\& \quad \times
 \delta\left(z-\frac{1-w_1-w_2+w_1 w_2 x}{1-w_1}\right) \df \Phi_{m}(k)
\,,\end{align}
where
\beq
\df \Phi_{m}(k)=(2\pi)^d \delta^d\left(k-\sum_{i=3}^{m+2}p_i\right) \prod_{i=3}^{m+2} \frac{\df^dp_i}{(2\pi)^d} (2\pi)\delta_+(p_i^2).
\eeq
The kinematic variables are defined in the following domains,
\bea
x\in[0,1],\hspace{1cm} w_1<0,\hspace{1cm}w_2>0,\hspace{1cm}q^2<0.
\eea
We can now express the desired partonic coefficient function defined in \eq{partdef}
in terms of the partonic coefficient function differential in the above variables,
\begin{align} \label{eq:partdef_2}
 \frac{\df\hat\eta_{i+h\rightarrow j+X}}{\df\myzeta\df\cO} &
 = \int\df w_1 \df w_2 \df x \, \delta[\myzeta - \myzeta(w_1, w_2, x)] \, \delta[\cO - \cO(w_1, w_2, x)] \,
   \frac{\df\hat\eta_{i+h\rightarrow j+X}}{\df w_1 \df w_2 \df x}
\,,\nn\\
 \frac{\df\hat\eta_{i+h\rightarrow j+X}}{\df w_1 \df w_2 \df x} &
 = \frac{1}{\hat\sigma_0} \frac{\cN_i}{2|q^2|} \sum\limits_{m=0}^\infty
   \int\!\frac{\df\Phi_{1+m}}{\df w_1 \df w_2 \df x} \, |\cM_{i+h\rightarrow j+m}|^2
\,.\end{align}
The second line is the central object in this work, from which all desired observables can be easily projected out.
It can be expanded as
\begin{align} \label{eq:eta_ij_1}
 &\frac{\df\hat\eta_{i+h\rightarrow j+X}}{\df w_1 \df w_2 \df x}
 = \sum_{\ell=0}^\infty \left(\frac{\as}{\pi}\right)^{\ell}
   \frac{\df\eta_{ij}^{(\ell)}}{\df w_1 \df w_2 \df x}
\\\nn&
 = \eta_{ij}^V \delta(w_1)\delta(w_2)\delta(x)
 \,+\, \sum_{\ell=1}^\infty \left(\frac{\as}{\pi}\right)^{\ell}
   \sum_{n,m=1}^\ell (-w_1)^{-1-m\eps} w_2^{-1-n\eps}
   \frac{\df\eta_{ij}^{(\ell,m,n)}(w_1,w_2,x,q^2)}{ \df w_1 \df w_2 \df x}
\,.\end{align}
Here, we have expanded $\hat\eta_{i+h\rightarrow j+X}$ in the strong coupling constant $\as/\pi$, and denote the coefficients as $\eta_{ij}^{\ell}$ for brevity.
In the second line we have split off the terms $\eta_{ij}^V$ which arise purely from Born contributions and virtual corrections.
The remaining functions $\eta_{ij}^{(\ell,m,n)}$ are separately holomorphic in the vicinity of $w_1=0$ and $w_2=0$.

The benefit of using the variables defined in \eq{supervars} is that together with $q^2$
they fully specify the momentum $p_2$ and thus are sufficient to express in $\hat\eta_{i+h\to j+X}$ differential in $p_2$.
For example, the Lorentz-invariant momentum fractions defined in \eq{zetadef} are given by
\begin{align} \label{eq:familiarvars}
 \myzeta &= \frac{1-w_2}{1-w_1-w_2+w_1 w_2 x}
\,,\qquad
 z = \frac{1-w_1-w_2+w_1 w_2 x}{1-w_1}
\,.\end{align}

\subsection{Transverse Momenta}
\label{sec:ptdef}
In SIDS, two particular definitions of transverse momentum play a key role. 
These two different definitions of transverse momentum are most naturally measured in two different inertial frames.
We define the \emph{infinite momentum frame} (also referred to as \emph{Breit frame}) and the \emph{hadron frame} as follows:
\beq
\label{eq:framedef}
\begin{array}{c|c}
\hspace{1cm}\text{Infinite Momentum Frame} \hspace{1cm}&\hspace{1cm} \text{Hadron Frame} \hspace{1cm}
\\
\hline
\\
q=(0,\vec 0,Q) & q=(q^0,\qt,q_z) \\ \\
P_1=E_1(1,\vec 0,1) & P_1=E_1(1,\vec 0,1) \\ \\
P_2=\Bigl(\sqrt{P_{2z}^2 + P_{2T}^2} , \vec P_{2T} ,P_{2z}\Bigr) & P_2=E_2(1,\vec 0,-1)
\end{array}
\,.\eeq
Here, $E_1$ and $E_2$ represent the energies of the initial and final state hadrons, respectively.
The explicit vectors in the above table are Euclidean vectors. 
The momentum component $|\qt|$ of the momentum $q$  is orthogonal to the plane spanned by the momenta $P_1$ and $P_2$ of the hadrons and is most naturally measured in the hadron frame.
The momentum component $|\vec{P}_{2T}|$ of the momentum $P_2$ is orthogonal to the plane spanned by the momenta $q$ and $P_1$ and is most naturally measured in the infinite momentum frame.
We express both transverse momenta in terms of Lorentz invariant quantities by
\bea \label{eq:pt_defs}
|\vec{P}_{2T}|^2&=&\frac{S^2 x_B^2}{Q^2}\left(1+\frac{Q^2}{S}\frac{x_F}{x_B}\right),\nonumber\\
|\qt|^2&=&Q^2\left(1+\frac{Q^2}{S}\frac{x_F}{x_B}\right).
\eea
Here,
\beq
S=(P_1+P_2)^2 = \frac{z}{x_B}\frac{x_F}{\myzeta} s
\eeq
is the invariant mass of the dihadron system.
Inserting the parametrisation in terms of $w_1$, $w_2$ and $x$ as defined in \eq{supervars},
the two transverse momenta of interest can be expressed in a Lorentz-invariant fashion as
\bea \label{eq:P2T}
|\vec{P}_{2T}|^2&=&x_F^2 \frac{q^2 w_1 w_2 (1-x)(1-w_1-w_2+w_1 w_2 x)}{(1-w_1)^2 (1-w_2)^2}
\,,\nn\\
|\qt|^2&=&\frac{q^2 w_1 w_2 (1-x)}{1-w_1-w_2+w_1 w_2 x}
\,.\eea
Below, we will be mostly interested in the limit that $|\vec{P}_{2T}|^2$, or equivalently $|\qt|^2$, becomes small.
We will approach this limit by considering the limit $w_2\to 0$, for which one obtains the simple relation
\beq \label{eq:framerels}
 \lim_{w_2 \to 0}: \qquad
|\vec{P}_{2T}|^2= x_F^2 |\qt|^2 = x_F^2 \frac{q^2 w_1 w_2 (1-x)}{1-w_1}
\,.\eeq

\section{Crossing from Production to DIS Cross Sections}
\label{sec:crossing}

In the previous section, we introduced the SIDIS process $P(P_1) + h(q) \to H(-P_2) + X(-k)$ for the scattering off an electroweak boson $h$ off the proton $P$,
thereby producing a detected final-state hadron $H$ in association with additional hadronic radiation.
The associated cross section is related by a factorization theorem to the partonic process $i(p_1) + h(q) \to j(-p_2) + X(-k)$,
which we describe by the partonic coefficient function $\eta_{ij}$ where we are fully differential in $p_1$ and $p_2$, but integrate over $k$.

We now want to relate, i.e.~\emph{cross}, this partonic configuration to the one where both partons are in the initial state and produce an outgoing electroweak boson $h$, which we hence refer to as ``production''.
Concretely, we study the crossing relation
\begin{align}
 p(p_1) + h(q) \rightarrow p(-p_2) + X(-k) \qquad\longleftrightarrow\qquad p(p_1)+ p(p_2) \rightarrow h(-q) + X(-k)
\,,\end{align}
where, as always, we choose all momenta as ingoing.

Recently, we have studied this production process \refscite{Ebert:2020lxs,Dulat:2017prg,Dulat:2017brz}.
In particular, in \refcite{Ebert:2020lxs} we showed that the corresponding partonic coefficient function is given by
\begin{align} 
\label{eq:sigma_part}
  \frac{\df \eta_{ij}^{\text{production}}}{ \df Q^2  \df \wa \df \wb \df  x}  &
 = \frac{1}{\hat \sigma_0} \frac{\cN_{ij}^{\text{production}}}{2 Q^2} \sum_{X_n}\int  \frac{\df\Phi_{h+n} }{ \df \wa \df \wb \df  x}\, |\cM_{ij\to h+X_n}|^2
\,,\end{align}
where the differential phase space for $h + n$ partons is given by
\begin{align} \label{eq:dPhi_wa_wb_x}
 \frac{\df\Phi_{h+m}}{\df \wa\df \wb \df x} &
 = \frac{\left(\frac{\wa\wb q^2}{1-\wa-\wb+\wa \wb x}\right)^{1-\epsilon} (1-x)^{-\epsilon}}{(4\pi)^{2-\epsilon} \Gamma(1-\epsilon)}
 \, \theta[x(1-x)] \, \theta(\wa) \, \theta(\wb) \, \df\Phi_m(k)
\,.\end{align}
Here, all variables are identical to the ones introduced in \sec{vardef} for SIDIS.
In particular, note that the squared matrix elements are identical in the DIS and production case up to the crossing of momenta $p_2$ and $q$. 
Furthermore, in both cases, production and DIS, the final state radiation is integrated over the phase space $\df \Phi_m$.
The dependence of the cross sections on the momenta $p_2$ and $q$ is fully retained.

In order to relate the partonic coefficient function of DIS to production, or vice versa, we need to understand the analytic structure of the partonic coefficient function.
Crossing $p_2$ and $q$ changes the sign of the numerical value of the invariants $s$ and $w_1$, and consequently it is important to understand the analytic branch structure of the partonic coefficient functions at $s=0$ and $w_1=0$.
Since $s$ is the only variable with explicit mass dimension in our choice of independent variables, it immediately follows that the partonic coefficient function at $\cO(\as^n)$ depends on $s$ only through the multiplicative factor $s^{-n \eps}$.
The analytic dependence on $w_1$ was already hinted at in \eq{eta_ij_1}, but needs to be investigated in more detail.

\begin{figure*}
 \centering
 \includegraphics[width=0.5\textwidth]{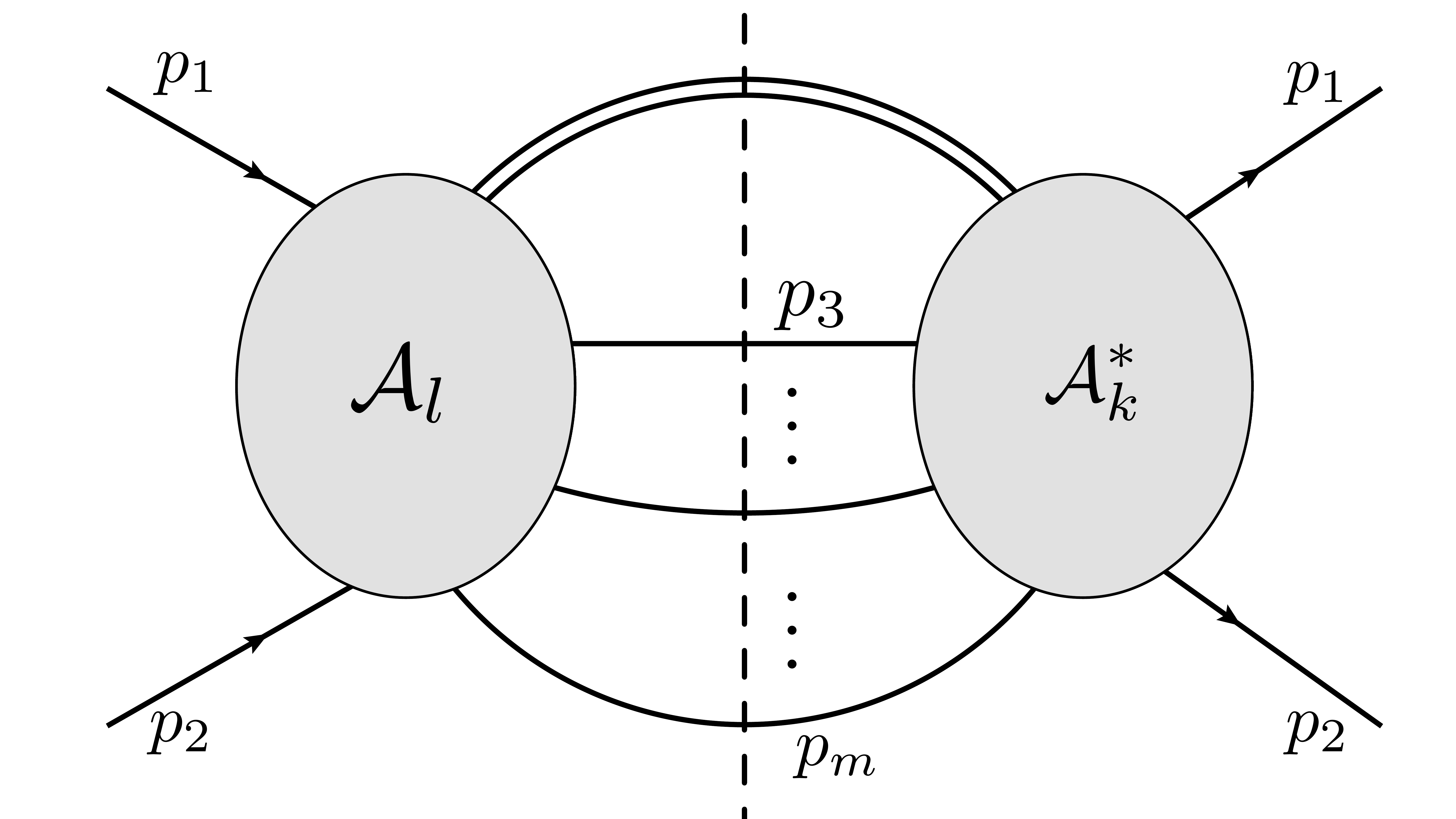}
 \caption{Schematic picture of the interference of a $l$-loop Feynman diagram with a complex-conjugate $k$-loop Feynman diagram.}
 \label{fig:Interfer}
\end{figure*}

The partonic coefficient function comprises of amplitudes interfering with complex conjugate amplitudes, integrated over the $m$-parton phase space.
This can be further split into interference of $l$-loop Feynman diagrams with conjugate $k$-loop Feynman diagrams, as illustrated in \fig{Interfer}.
Similar to the decomposition of the partonic coefficient function in \eq{eta_ij_1}, the analytic structure of the depicted interference diagram can be decomposed as
\begin{align} \label{eq:amplintef}
 & \int\!\df\Phi_m \, \Re(\cA_l \cA_k^*)
 = (s w_1 w_2)^{-m \epsilon} \times \Biggl\{ \sum\limits_{i_1,i_2=0}^{l}\sum\limits_{j_1,j_2=0}^{k} f^{(i_1,i_2,j_1,j_2)}(w_1,w_2,x)
\nn\\& \quad \times
   \Re\Bigl\{ \Bigl[ (-s)^{(i_1+i_2-l)\epsilon}(s w_1)^{-i_1 \epsilon}  (sw_2)^{-i_2 \epsilon} \Bigl]
              \Bigl[ (-s)^{(j_1+j_2-k)\epsilon}(s w_1)^{-j_1 \epsilon}  (sw_2)^{-j_2 \epsilon} \Bigr]^* \Bigr\} \Biggr\}
\,.\end{align}
Here, the functions $f^{(i_1,i_2,j_1,j_2)}(w_1,w_2,x)$ do not contain any branch cuts at $s=0$, $w_1=0$ or $w_2=0$.
When performing a computation of analytic partonic coefficient functions, it is easy and often useful to keep track of the individual functions $f^{(i_1,i_2,j_1,j_2)}(w_1,w_2,x)$.
The second line in \eq{amplintef} differs between DIS and production kinematics due to the different signs of $s$ and $w_1$.
Explicit phases occur in a given loop amplitude depending on the kinematic configuration of the external momenta.
The phases are easily determined by equipping the Lorentz-invariant scalar products $s$, $s w_1$ and $s w_2$ with a definite Feynman prescription,
\beq
(p_i+p_j)^2 \quad\rightarrow\quad (p_i+p_j)^2+ i0.
\eeq
Crossing from DIS to production kinematics then requires us to analytically continue the second line of \eq{amplintef}.
As an example, we consider the case $i_1=i_2=k=j_1=j_2=0$,
\beq
\underbrace{\Re\left[(-s-i0)^{-l \epsilon}\right]}_{\text{DIS}}
\qquad\longleftrightarrow\qquad
\underbrace{\cos(l \pi\epsilon) \, \Re\left[(s+i0)^{-l\epsilon}\right]}_{\text{production}}
\,.\eeq
The same analytic structure as outlined above for the interference of two Feynman diagrams naturally holds for the entire partonic coefficient function as well,
\begin{align} \label{eq:PCFanalytic}
 &\frac{\df\eta_{ij}^{(m+l+k)}}{\df Q^2 \df w_1 \df w_2 \df x}
 = (s w_1 w_2)^{-m \epsilon} \times \Biggl\{ \sum\limits_{i_1,i_2=0}^{l}\sum\limits_{j_1,j_2=0}^{k} \frac{\df\eta_{ij}^{(m+l+k,i_1,i_2,j_1,j_2)}}{\df Q^2 \df w_1 \df w_2 \df x}
\nn\\& \quad \times
 \Re \Bigl\{ \Bigl[ (-s)^{(i_1+i_2-l)\epsilon}(s w_1)^{-i_1 \epsilon}  (sw_2)^{-i_2 \epsilon} \Bigr]
             \Bigl[ (-s)^{(j_1+j_2-k)\epsilon}(s w_1)^{-j_1 \epsilon}  (sw_2)^{-j_2 \epsilon} \Bigr]^* \Bigr\} \Biggr\}
\,.\end{align}
Once the universal functions $\eta_{ij}^{(m+l+k,i_1,i_2,j_1,j_2)}$ are identified, it is easy to perform the analytic continuation between DIS and production kinematics.
The above was observed and explicitly verified for the computation of the ingredients of Higgs and DY production up to N$^3$LO
in \refscite{Dulat:2017brz,Anastasiou:2013mca,Dulat:2014mda,Anastasiou:2015yha,Anastasiou:2013srw}
and holds in particular for the interference of amplitudes for massless QCD corrections for the processes under consideration.
We note that it is of course also possible to relate DIS or production kinematics to partonic cross sections
where only the electroweak gauge boson is in the initial state and all partons are in the final state, for example $e^+\,e^-$ annihilation.

In addition to the analytic continuation from DIS to production kinematics there are some other, trivial differences in the partonic coefficient functions.
First, the overall normalisation factor $\cN_i$ and $\cN_{ij}^{\text{production}}$ differ, which can be trivially accounted for.
Second, the phase space measure $\df\Phi_{1+m}$ and $\df\Phi_{h_+m}^{\text{production}}$ differ by factors depending on the kinematic variables.
However, this difference is accounted for by a simple multiplicative factor that does not require any additional analytic continuation.
With this we have identified all differences between DIS and production kinematics in bare, partonic coefficient functions and can relate one to the other
as long as the required analytic information is retained in the computation of one of them.

Analytic continuation of processes and universal anomalous dimensions, such as splitting functions appearing in the evolution of parton densities and fragmentation function have a long history~\cite{Drell:1969jm,Stratmann:1996hn,Blumlein:2000wh,Muller:2012yq,Mitov:2006ic,Moch:2007tx,Almasy:2011eq,Dokshitzer:2005bf,Basso:2006nk,Neill:2020bwv,Chen:2020uvt,Kologlu:2019mfz,Korchemsky:2019nzm,Dixon:2019uzg,Chen:2020vvp}.
The fact that our setup is differential in all four momenta that are crossed from one kinematic configuration to another allows us to frame crossing purely in terms of analytic continuation.

\section{Collinear limit of partonic coefficient functions}\label{sec:collinear_limit}
In this section we briefly review the method introduced in \refcite{Ebert:2020lxs} to expand cross sections in the kinematic limit
where all final-state radiation becomes collinear to the parton with momentum $p_1$ or $p_2$.
In order to illustrate this, it is instructive to decompose the momentum $k$ into its components along these directions,
\beq
k^\mu=p_1^\mu k_1+p_2^\mu k_2 +k_\perp^\mu.
\eeq
Here, the $k_\perp$ component is chosen orthogonal to $p_1$ and $p_2$.
In order to illustrate the collinear limit with respect to either massless parton we introduce an auxiliary rescaling parameter $\lambda$ and indicate the collinear limit by
\bea \label{eq:coll_trafo}
 p_1{-}\mathrm{collinear}: \qquad k^\mu &\rightarrow&  \quad p_1^\mu k_1+\lambda^2 p_2^\mu k_2 +\lambda k_\perp^\mu
\,,\nn\\
 p_2{-}\mathrm{collinear}: \qquad k^\mu &\rightarrow&  \lambda^2 p_1^\mu k_1 + \quad p_2^\mu k_2 +\lambda k_\perp^\mu
\,.\eea
The respective limit is then achieved by taking $\lambda \to 0$.
The variables $w_1$, $w_2$ and $x$ defined in \eq{supervars} were chosen such that the action of either collinear rescaling transformation in \eq{coll_trafo} on the partonic coefficient function simply amounts to a rescaling of $w_{1,2}$.
Specifically, in the $p_1$-collinear limit only $w_1$ is rescaled, while in the $p_2$-collinear limit only $w_2$ is rescaled, while the other variables are not affected, 
\bea
 p_1{-}\mathrm{collinear}: \qquad w_1 \to \lambda^2 w_1 \,,\quad w_2 \to w_2  \,,\quad\quad x \to x
\,,\nn\\
 p_2{-}\mathrm{collinear}: \qquad w_1 \to w_1 \,,\quad\quad w_2 \to \lambda^2  w_2  \,,\quad x \to x
\,.\eea
An expansion of our partonic coefficient function in the $p_{1,2}$-collinear limit is thus equivalent to an expansion in $w_{1,2}$.
More details on how such an expansion can be performed for multi-loop partonic coefficient functions can be found in \refcite{Ebert:2020lxs}.

A key difference between the $p_1$- and $p_2$-collinear limit is that the former corresponds to a collinear initial-state singularity,
which were already discussed in \refscite{Ebert:2020lxs,Ebert:2020unb,Ebert:2020yqt}, while the latter corresponds to collinear final-state singularity.
Here, we only only briefly look at the impact of the $p_1$-collinear limit on the more familiar variables given in \eqs{familiarvars}{pt_defs},
\begin{align}
 p_1{-}\mathrm{collinear}: \qquad
 \frac{|\vec{P}_{2T}|^2}{x_F^2} = |\qt|^2 = \frac{q^2 w_1 w_2 (1-x)}{1-w_2}
\,,\quad
 \myzeta \to 1
\,,\quad
 z \to 1-w_2
\,.\end{align}
Note, that the $p_1$-collinear limit of the phase space is identical for DIS and production kinematics up to the domain of the variables,
\beq
\lim\limits_{p_1-\rm coll} \frac{\df \Phi_{1+m}}{\df w_1 \df w_2 \df x}\sim\lim\limits_{p_1-\rm coll} \frac{\df\Phi_{h+m}}{\df w_1\df w_2\df x} \,.
\eeq
Furthermore, in the strict $p_1$-collinear limit, which is defined by only retaining momentum modes in loop integrals where the loop momentum itself is collinear to $p_1$~\cite{Ebert:2020lxs},
none of the partonic coefficient functions require any analytic continuation when crossing between DIS and production kinematics.
Thus, up to overall normalization factors the strict $p_1$-collinear limit agrees between production and DIS kinematics.
Of course, this is an immediate consequence of the universality of collinear dynamics of QCD and the factorization of collinear initial-state singularities.

The limit of all final-state radiation becoming collinear to the momentum $p_2$ corresponds to collinear final-state singularities,
which were not discussed in \refcite{Ebert:2020lxs} and are the main focus of this article.
In this limit, the familiar variables in \eqs{familiarvars}{pt_defs} become
\begin{align} \label{eq:supervars_p2}
 p_2{-}\mathrm{collinear}: \qquad
 \frac{|\vec{P}_{2T}|^2}{x_F^2} = |\qt|^2 = \frac{q^2 w_1 w_2 (1-x)}{1-w_1}
\,,\quad
 \myzeta \to \frac{1}{1-w_1}
\,,\quad
 z \to 1
\,.\end{align}
Note, that $\myzeta$ in the $p_2$-collinear limit behaves reciprocal to $z$ in the $p_1$-collinear limit, which is a consequence of their definition in eq.~\eqref{eq:zetadef}.
In contrast to the $p_1$-collinear limit, in the $p_2$-collinear limit the phase space for DIS and production kinematics differ slightly by
\beq
\lim\limits_{p_2-\rm coll} \frac{\df \Phi_{1+m}}{\df w_1 \df w_2 \df x }\sim(1-w_1)^{-3+2\epsilon}\lim\limits_{p_2-\rm coll} \frac{\df\Phi_{h+m}}{\df w_1\df w_2\df x} \,.
\eeq
Furthermore, in order to cross from production to DIS kinematics it is necessary to analytically continue parts of the partonic coefficient function, as outlined in \sec{crossing}.

\section{Calculation of the TMD Fragmentation Functions}
\label{sec:TMDFFExt}

In this section we calculate the TMDFFs at N$^3$LO from a perturbative calculation of the SIDIS process $P(P_1) + h(q) \rightarrow H(-P_2) + X(-k)$.
We will briefly review the required factorization for SIDIS in the limit of small transverse momentum in \sec{SIDIS_factorization},
before showing in \sec{TMDFF_calculation} how it relates to the kinematic limit where the final-state momenta $P_2$ and $k$ are collinear to each other.
In \sec{results}, we discuss our results for the TMDFFs.

In this section, it will be useful to introduce lightcone coordinates,
which we define in terms of two lightlike reference vectors
\begin{align} \label{eq:n_bn}
 n^\mu = (1, 0, 0, 1) \,,\qquad \bn^\mu = (1, 0, 0, -1)
\,,\end{align}
which obey $n^2 = \bn^2 = 0$ and $n{\cdot}\bn = 2$.
Any four momentum $p^\mu$ can then be decomposed as
\beq
 p^\mu = p^- \frac{n^\mu}{2} + p^+ \frac{\bn^\mu}{2} + p_\perp^\mu \equiv (p^+, p^-, p_\perp)
\,,\eeq
where $p^- \equiv \bn \cdot p$ and $p^+ \equiv n \cdot p$.
We will always denote transverse vectors in Minkowski space as $p_\perp^\mu = (0, \pt, 0)$ where $\pt$ is a Euclidean two vector, such that $p_\perp^2 = -\pt^2 \equiv - p_T^2$.

\subsection{SIDIS factorization at small transverse momentum}
\label{sec:SIDIS_factorization}

We consider the unpolarized SIDIS process in \eq{DIS} in a frame where the incoming proton $P$ and outgoing hadron $H$ are aligned along the lightcone vectors in \eq{n_bn}, i.e.
\begin{align} \label{eq:hadron_frame_coordinates}
 P_1^\mu = P_1^- \frac{n^\mu}{2} \,,\qquad P_2^\mu = P_2^+ \frac{\bn^\mu}{2}
\,.\end{align}
In this frame, the momentum $q^\mu$ of the electroweak boson $h$ is given by
\begin{align}
 q^\mu = (q^+, q^-, q_\perp) \qquad\text{with}\qquad -Q^2 = q^2 = q^+ q^- - q_T^2
\,.\end{align}
In particular, it has a nonvanishing transverse momentum $\qt$.
Note, that the above coordinates correspond to the hadron frame introduced in sec.~\ref{sec:ptdef}.

The factorization of the SIDIS cross section in the limit of small transverse momentum, $q_T \ll Q$, was first derived in~\cite{Ji:2004wu} and elaborated on in \refscite{Aybat:2011zv, Collins:1350496}.
We follow the notation established in the treatment of TMD factorization within Soft-Collinear Effective Theory (SCET)~\cite{Bauer:2000ew, Bauer:2000yr, Bauer:2001ct, Bauer:2001yt} in the formalism of the rapidity renormalization group equation~\cite{Chiu:2012ir, Li:2016axz}.
For Drell-Yan like processes, the factorized cross section is given by%
\begin{align} \label{eq:xs_SIDIS_q}
 \frac{\df\sigma}{\df x_F \, \df^2 \qt} &
 = \hat\sigma_0 \, x_F^2 \, \sum_{i,j} H_{ij}(q^2, \mu) \!\int\!\frac{\df^2\bt}{(2\pi)^2} e^{\img \qt \cdot \bt}
   \tilde B_i\Bigl(x_B, \bt, \mu, \frac{\nu}{\omega_a}\Bigr) \tilde D_{H/j}\Bigl(x_F, \bt, \mu, \frac{\nu}{\omega_b}\Bigr)
   \nn\\&\qquad  \times
   \tilde S_q(b_T, \mu, \nu)
   \times \Bigl[1 + \cO\Bigl(q_T^2/Q^2\Bigr)\Bigr]
\,,\end{align}
see \app{factorization} for more details.
In \eq{xs_SIDIS_q}, $\hat\sigma_0$ is the same Born cross section as before, the sum runs over all parton flavors $i,j$ contributing to the Born process $i + h \to j$, and the hard function $H_{ij}$ encodes virtual corrections to the Born process.
As is common, the factorization in \eq{xs_SIDIS_q} is expressed in Fourier space, with $\bt$ Fourier conjugate to $\qt$.
The TMD beam and fragmentation functions $\tilde B_i(x_B, \bt)$ and $\tilde D_{H/j}(x_F,\bt)$ encode the effect of radiation collinear
to the incoming proton and outgoing hadron, respectively, and are defined below.
They depend on $\bt$ and the momentum fractions $x_{B,F}$ as defined in \eq{xB_xi}.
The soft function $\tilde S_q(b_T)$ encodes the transverse recoil due to soft radiation, and is independent of the quark flavors $i$ and $j$.
\Eq{xs_SIDIS_q} depends not only on the common renormalization scale $\mu$, which we take as usual as the $\overline{\text{MS}}$ scale,
but also on the scale $\nu$ that arises from the regularization of so-called rapidity divergences~\cite{Collins:1981uk,Collins:1350496,Becher:2010tm,Becher:2011dz,GarciaEchevarria:2011rb,Chiu:2011qc,Chiu:2012ir,Li:2016axz,Rothstein:2016bsq,Ebert:2018gsn},
for which we employ the exponential regulator of \refcite{Li:2016axz}.
The momentum fractions $\omega_{a,b}$ in \eq{xs_SIDIS_q} are defined as the lightcone components
\begin{align} \label{eq:omegadef}
 \omega_a = x_B P_1^- \,, \quad \omega_b = -\frac{P_2^+}{x_F} \,, \qquad\Rightarrow\qquad \omega_a \omega_b \approx Q^2
\,.\end{align}
They are closely related to the Collins-Soper scale $\myzeta_{a,b} \propto \omega_{a,b}^2$~\cite{Collins:1981uk,Collins:1981va}.

For gluon-induced processes, the factorized cross section reads
\begin{align} \label{eq:xs_SIDIS_g}
 \frac{\df\sigma}{\df x_F \, \df^2 \qt} &
 = \hat\sigma_0 \, x_F^2 \,2 H_{\rho\sigma\rho'\sigma'}(q^2, \mu) \!\int\!\frac{\df^2\bt}{(2\pi)^2} e^{\img \qt \cdot \bt}
   \tilde B_g^{\rho\sigma}\Bigl(x_B, \bt, \mu, \frac{\nu}{Q}\Bigr) \tilde D_{H/g}^{\rho'\sigma'}\Bigl(x_F, \bt, \mu, \frac{\nu}{Q}\Bigr)
 \nn\\&\qquad\times
   \tilde S_g(b_T, \mu, \nu)
   \times \Bigl[1 + \cO\Bigl(q_T^2/Q^2\Bigr)\Bigr]
\,.\end{align}
The only difference to \eq{xs_SIDIS_g} is the Lorentz structure of $\tilde B_g^{\rho\sigma}$ and $\tilde D_{H/g}^{\rho\sigma}$,
which arises due to the helicity structure of the gluon field,
One can decompose the gluon TMDFF as
\begin{align} \label{eq:D_decomp}
 \tilde D_{H/g}^{\rho\sigma}(x_F, \bt) = \frac{g_\perp^{\rho\sigma}}{2} \tilde D_{H/g}(x_F, \bt) + \Bigl(\frac{g_\perp^{\rho\sigma}}{2} + \frac{b_\perp^\mu b_\perp^\nu}{b_T^2}\Bigr) \tilde D'_{H/g}(x_F, \bt)
\,,\end{align}
where we suppressed the scales for brevity. The decomposition of $\tilde B_g^{\rho\sigma}$ has the same structure as \eq{D_decomp}.
We will only consider Higgs production, where due to the scalar nature of the Higgs boson
\begin{align}
 H^{\rho\sigma\rho'\sigma'}(q^2, \mu) = H(q^2, \mu) g_\perp^{\rho\rho'} g_\perp^{\sigma\sigma'}
\,,\end{align}
and thus we only require the combination
\begin{align}
 2 H_{\rho\sigma\rho'\sigma'} \tilde B_g^{\rho\sigma} \tilde D_{H/g}^{\rho'\sigma'} &
 = H \bigl( \tilde B_g \tilde D_{H/g} + \tilde B'_g \tilde D'_{H/g} \bigr)
\,,\end{align}
where we suppressed all arguments for brevity.
Since this structure is very similar to the combination in \eq{xs_SIDIS_q}, in the following we will always use the form in \eq{xs_SIDIS_q},
with the implicit understanding that the $B'_g D'_g$ term has to be added for Higgs production.
Furthermore, since $B'_g = \cO(\as)$ and $D'_g = \cO(\as)$, their NNLO results are sufficient to describe Higgs production at N$^3$LO.
They have already been calculated in \refcite{Luo:2019bmw}, and we will not consider them in our calculation of the N$^3$LO TMDFFs.

Before proceeding, we remark that the precise form of the beam, fragmentation and soft functions in \eqs{xs_SIDIS_q}{xs_SIDIS_g}
depends on the chosen rapidity regulator. In our work, we will use the exponential regulator of \refcite{Li:2016axz},
and the ensuing rapidity renormalization scale is denoted as $\nu$, but many other rapidity regularization schemes are known in the literature~\cite{Collins:1350496,Becher:2010tm,Becher:2011dz,GarciaEchevarria:2011rb,Chiu:2012ir,Ebert:2018gsn}.
The scheme ambiguity can be eliminated by combining the beam and fragmentation functions with the soft function,
which in our case reads
\begin{align} \label{eq:B_to_f}
 \tilde f_i^\TMD(x_B, \bt, \mu, \zeta_a) &= \tilde B_i(x_B, \bt, \mu, \frac{\nu}{\omega_a}) \sqrt{S(b_T, \mu, \nu)}
\,,\nn\\
 \tilde D_{H/j}^\TMD\Bigl(x_F, \bt, \mu, \zeta_b \Bigr) &= \tilde D_{H/j}\Bigl(x_F, \bt, \mu, \frac{\nu}{\omega_b} \Bigr) \sqrt{S(b_T, \mu, \nu)}
\,.\end{align}
These combinations are manifestly $\nu$ independent, reflecting the independence of the rapidity regulator.
As is common, we have introduced the so-called Collins-Soper scale $\zeta_{a,b} = \omega_{a,b}^2$, a remnant of the rapidity regularization.
Note that while there is an established notation distinguishing TMD beam functions $\tilde B_i$ and TMDPDFs $\tilde f_i$,
so far no such notation exists for the TMDFF. To make clear which function we refer to, we will label the TMDFF including the soft function by an explicit superscript ``TMD''.

Similar combinations as in \eq{B_to_f} can be constructed in all viable rapidity regulators (often, these functions are combined at the bare level prior to renormalization, with UV renormalization applied to the product).
To calculate the TMDFF itself, one has to calculate collinear and soft matrix elements separately,
and hence it is natural to separate the fragmentation functions from the soft function.
Thus, we will provide results both for the scheme-dependent $\tilde D_{H/j}$ before and the scheme-independent  $\tilde D^\TMD_{H/j}$ after combination with the soft function.

The TMDFFs in \eqs{xs_SIDIS_q}{xs_SIDIS_g} are well-defined QCD hadronic matrix elements~\cite{Mulders:2000sh}.
Using SCET notation, the bare fragmentation functions are defined as
\begin{align} \label{eq:def_D}
 \tilde D_{H/q} (x_F, \bt, \eps, \tau ) &
 = \frac{1}{4 N_c} \frac{1}{x_F} \sum_X \int\!\frac{\df b^-}{4\pi} e^{\img P^+ b^- / (2 x_F)} {\rm Tr} \braket{0 | \slashed{n} \, \chi_\bn(b) | H X} \braket{H X | \bar \chi_\bn(0) | 0 }
\,,\nn\\
 \tilde D^{\mu\nu}_{H/g} (x_F,\bt, \eps, \tau ) &
 = -\frac{P^+}{x_F^2} \sum_X \int\frac{\df b^-}{4\pi} e^{\img P^+ b^- /(2x_F)}
   \braket{0 | \cB_{\bn\perp}^\mu(b) | H X} \braket{H X | \cB_{\bn\perp}^\nu(0) | 0}
\,.\end{align}
Here, we make explicit that we regulate UV divergences by working in $d=4-2\eps$ dimensions and regulate rapidity divergences using the exponential regulator of \refcite{Li:2016axz}.
In \eq{def_D}, the sum is over all additional hadronic final states $X$, the trace is over color and spin, and $P$ is the momentum of the hadron $H$.
The fields $\chi_\bn$ and $\cB_{\bn\perp}^\mu$ are collinear quark and gluon fields in SCET, with the pair of fields in each equation separated by $b^\mu = (0, b^-, b_\perp)$.
The matrix elements in \eq{def_D} are defined in the hadron frame as specified in \eq{hadron_frame_coordinates},
i.e.~the outgoing hadron $H$ defines the lightcone direction $\bn^\mu$, and $\bt$ is transverse to it.
\\\indent
For perturbative $b_T \gtrsim \lqcd^{-1}$, the TMDFF can be matched perturbatively onto the collinear FF.
For the renormalized TMDFF, this relation reads~\cite{Aybat:2011zv, Collins:1350496}
\begin{align} \label{eq:matching_D}
 \tilde D_{H/j}\Bigl(x_F, \bt, \mu, \frac{\nu}{\omega_b}\Bigr) &
 = \sum_{j'} \int_{x_F}^1 \frac{\df z}{z^3} d_{h/j'}(z, \mu) \, \tilde\cC_{jj'}\Bigl(\frac{x_F}{z}, \bt, \mu, \frac{\nu}{Q}\Bigr)
\nn\\&
 = \sum_{j'} \int_{x_F}^1 \frac{\df z}{z} d_{h/j'}\Bigl(\frac{x_F}{z}, \mu\Bigr) \, \frac{z^2}{x_F^2}  \tilde\cC_{jj'}(z, \bt, \mu, \frac{\nu}{Q})
\,,\end{align}
where the matching coefficients $\tilde\cC_{jj'}$ are perturbatively calculable.
In the second line in \eq{matching_D} we have replaced $z \to x_F/z$, which will be more convenient for our extraction of $\tilde\cC_{jj'}$.
Corrections to \eq{matching_D} are suppressed as $\cO(b_T \lqcd)$.
With this we may rewrite the TMDFF of eq.~\eqref{eq:matching_D} in terms of the Mellin convolution
\bea  \label{eq:matching_D_2}
x_F^2 \tilde D_{H/j}\Bigl(x_F, \bt, \mu, \frac{\nu}{\omega_b}\Bigr) &=&  \sum_{j'}  \tilde{\cK}_{jj^\prime }\Bigl(x_F,\vec b_T,\mu,\frac{\nu}{\omega_b} \Bigr)  \otimes_{x_F}  d_{h/j'}(x_F,\mu)\nonumber\\
 &=&  \sum_{j'}  \left[ x_F^2 \tilde{\cC}_{jj^\prime }\Bigl(x_F,\vec b_T,\mu,\frac{\nu}{\omega_b} \Bigr)\right] \otimes_{x_F}  d_{h/j'}(x_F,\mu).
\eea
Above, we implicitly defined the perturbative matching kernel
\beq
\label{eq:our_dearest_kernel}
 \tilde{\cK}_{jj^\prime }\Bigl(\myzeta,\vec b_T,\mu,\frac{\nu}{\omega_b} \Bigr) =\myzeta^2 \tilde{\cC}_{jj^\prime }\Bigl(\myzeta,\vec b_T,\mu,\frac{\nu}{\omega_b} \Bigr).
\eeq

The TMDFFs in \eq{def_D} are defined in a coordinate system where $P_2^\mu = P_2^+ \bn^\mu/2$ defines the lightcone direction and has vanishing transverse momentum, and hence $b_\perp$ is Fourier-conjugate to the transverse momentum of the parton that initiates the fragmentation process.
Alternatively we may consider the transverse momentum of the final state hadron $P_{2T}$ which is naturally defined in the infinite momentum frame, see sec.~\ref{sec:ptdef}, and following \refcite{Luo:2019hmp} we denotes this definition of the TMDFF $\tilde\cF_{H/j}$.
Since the two transverse momenta in these two frames are related as $\vec P_{T2} = - x_F \qt$, see \eq{framerels}, the two TMDFFs are related by
\begin{align} \label{eq:relation_D_F}
  \cF_{H/j}(x_F, \vec P_{2\perp}) &= D_{H/j}(x_F, - \vec P_{2\perp} / x_F)
\,,\nn\\
  \tilde \cF_{H/j}(x_F, \bt/x_F) &= x_F^{d-2} \tilde D_{H/j}(x_F, -\bt)
\,.\end{align}
The first relation is an immediate consequence of eq.~\eqref{eq:framerels}.
The second equation immediately follows upon Fourier transform in $d-2$ dimensions.

In \refcite{Luo:2019hmp}, the matching relation for the $\tilde\cF_{H/j}$ was written as
\begin{align}
 \tilde\cF_{H/j}\Bigl(x_F, \frac{\bt}{x_F}, \mu, \frac{\nu}{\omega}\Bigr) &
 = \sum_{j'} \int_{x_F}^1 \frac{\df \myzeta}{\myzeta} d_{H/j}\Bigl(\frac{x_F}{\myzeta}\Bigr) \tilde\cC_{j'j}^\text{\cite{Luo:2019hmp}}\Bigl(\myzeta, \frac{\bt}{\myzeta}, \mu, \frac{\nu}{\omega}\Bigr)
\,,\end{align}
and thus our kernels $\tilde\cK$ are identical to their kernels with rescaled arguments,
\begin{align}
 \tilde{\cK}_{jj^\prime }\Bigl(\myzeta,\vec b_T,\mu,\frac{\nu}{\omega_b} \Bigr) = \tilde\cC_{j'j}^\text{\cite{Luo:2019hmp}}\Bigl(\myzeta, \frac{\bt}{\myzeta}, \mu, \frac{\nu}{\omega}\Bigr)
\,.\end{align}

\subsection{TMD fragmentation functions from the collinear limit}
\label{sec:TMDFF_calculation}

The TMDFF can be obtained from the collinear limit of SIDIS following the same strategy applied in \refscite{Ebert:2020lxs, Ebert:2020yqt} to calculate the TMDPDF from the collinear limit of proton-proton scattering.
We start from the cross section differential in the transverse momentum $\qt$, which in the limit of small $q_T \ll Q$ is given by the factorization theorem in \eq{xs_SIDIS_q},
\begin{align} \label{eq:xs_SIDIS_q_2}
 \frac{\df\sigma}{\df x_F \, \df^2 \vec q_{\perp}} &
 = \hat\sigma_0  x_F^2\, \sum_{i,j} H_{ij}(q^2, \mu) \!\int\!\frac{\df^2\bt}{(2\pi)^2} e^{-\img \bt \cdot \qt}
   \tilde B_i\Bigl(x_B, \bt, \mu, \frac{\nu}{\omega_a}\Bigr) \tilde D_{H/j}\Bigl(x_F, \bt, \mu, \frac{\nu}{\omega_b}\Bigr)
   \nn\\&\qquad  \times
   \tilde S_q(b_T, \mu, \nu)
   \times \Bigl[1 + \cO\Bigl(q_T^2/Q^2\Bigr)\Bigr]
\,,\end{align}
The key insight is that the hard, beam, fragmentation and soft function in \eq{xs_SIDIS_q_2} encode different dynamics.
The hard function $H$ arises from hard virtual corrections to the Born process,
while $\tilde B$, $\tilde D$ and $\tilde S$ are constructed such that they only arise from $p_1$-collinear, $p_2$-collinear and soft momenta in loop integral and real emissions, respectively.
It follows that by calculating the strict $p_2$-collinear limit, defined such that both loop and real momenta are expanded in the $p_2$-collinear limit,
only the fragmentation function contributes to \eq{xs_SIDIS_q_2},
\begin{align} \label{eq:xs_SIDIS_coll}
 \bnlim
 \frac{\df\sigma}{\df x_F \, \df^{d-2} \qt} &
 = \hat\sigma_0 \, x_F^2 \, \sum_{i} \!\int\!\frac{\df^{d-2}\bt}{(2\pi)^{d-2}} e^{\img \bt \cdot \qt}
   f_i(x_B) \tilde D_{H/\bar i}(x_F, \bt)
\nn\\&
 = \hat\sigma_0 \, x_F^2 \, \sum_{i} f_i(x_B) D_{H/\bar i}\Bigl(x_F, \qt \Bigr)
\,.\end{align}
Here we used that the hard and soft functions are normalized to unity at tree level,
while the TMD beam function reduces to the PDF itself.
Note that \eq{xs_SIDIS_coll} is to be understood at the bare level, as only combining it with all other limits
will cancel all appearing infrared divergences, and thus $\qt$ and $\bt$ are treated in $d-2$ dimensions.
We also used that both photon and Higgs exchange are flavor diagonal to fix $j = \bar i$.

We want to relate \eq{xs_SIDIS_coll} to the SIDIS cross section defined in collinear factorization.
Combining \eqs{hadr2}{partdef_2}, we obtain
\begin{align} \label{eq:xs_SIDIS_fact}
 \frac{\df\sigma_{P+h \rightarrow H+X}}{\df {x_F}\df \qpsq} &
 = \hat\sigma_0 \sum_{i,j} \int_{x_B}^1\frac{dz}{z} \int_{x_F}^1\frac{\df\myzeta}{\myzeta}
   f_i\Bigl(\frac{x_B}{z}\Bigr) d_{H/j}\Bigl(\frac{x_F}{\myzeta}\Bigr)
   \\\nn&\quad\times
   \int\df w_1 \df w_2 \df x \,
   \frac{\df\hat\eta_{i+h\rightarrow j+X}}{\df w_1 \df w_2 \df x}
   \delta[\myzeta - \myzeta(w_1, w_2, x)] \, \delta[\qpsq - \qpsq(w_1, w_2, x)] \,
\,,\end{align}
where the expressions for $\myzeta$ and $\qpsq$ are given by \eqs{familiarvars}{P2T}.
In the limit that all final state radiation becomes collinear to $P_2$,
i.e.~$w_2 \to 0$,  all required variables becomes
\begin{align}
 p_2{-}\mathrm{collinear}: \qquad
 \qpsq \to  q^2 \frac{w_1 w_2 (1-x)}{1-w_1}
\,,\qquad
 \myzeta \to \frac{1}{1-w_1}
\,,\qquad
 z \to 1
\,.\end{align}
Note that in this limit, the partonic coefficient function scales as $\delta(1-z)$, see \eq{phi_1m}, and thus renders the convolution in $z$ trivial.
Furthermore, we fix $w_1 = -(1-\myzeta)/\myzeta$, and obtain
\begin{align} \label{eq:xs_SIDIS_fact_2}
 \bnlim \frac{\df\sigma_{P+h \rightarrow H+X}}{\df {x_F}\df \qpsq} &
 = \hat\sigma_0 \sum_{i,j} f_i(x_B)  \int_{x_F}^1\frac{\df\myzeta}{\myzeta^3}
   d_{H/j}\Bigl(\frac{x_F}{\myzeta}\Bigr)
   \\\nn&\quad\times
   \int\! \df w_2 \df x \, \delta\bigl[\qpsq -Q^2(1- \myzeta) w_2 (1-x) \bigr]
   \bnlim \frac{\df\hat\eta_{i+h\rightarrow  j+X}}{\df w_1 \df w_2 \df x}
\,.\end{align}
Comparing \eqs{xs_SIDIS_coll}{xs_SIDIS_fact_2}, we can immediately read off the relation between the perturbative matching kernel and take the Fourier transform with respect to $\qt$,
\begin{align}
 \tilde D_{H/ i}(x_F, \bt) &
 = \frac{1}{x_F^2} \sum_{j} \int_{x_F}^1\frac{\df\myzeta}{\myzeta} d_{H/j}\Bigl(\frac{x_F}{\myzeta}\Bigr)
   \int\frac{\df^{d-2}\qt}{\Omega_{d-3}(\qpsq)^{d/2-2}/2} e^{-\img \bt \cdot \qt}
   \\\nn&\quad\times
   \int \df w_2 \df x \, \delta\bigl[\qpsq-  Q^2 (1-\myzeta) w_2 (1-x) \bigr]
   \bnlim \frac{1}{\myzeta^2}\frac{\df\hat\eta_{\bar i+h\rightarrow  j+X}}{\df w_1 \df w_2 \df x}
\,.\end{align}
The perturbative matching kernel as defined in eqs.~\eqref{eq:matching_D} -- \eqref{eq:our_dearest_kernel} is then given by
\bea
\label{eq:C_naive}
\tilde{ \cK}^{\text{naive}}_{jj^\prime}(\myzeta,\bt)&=& \int\frac{\df^{d-2}\qt}{\Omega_{d-3}(\qpsq)^{d/2-2}/2}  e^{-\img \bt \cdot \qt}\\
&\times&
   \int \df w_2 \df x \,
   \delta\bigl[\qpsq-  Q^2 (1-\myzeta) w_2 (1-x) \bigr]
   \bnlim \frac{1}{\myzeta^2}\frac{\df\hat\eta_{\bar j+h\rightarrow  j^\prime+X}}{\df w_1 \df w_2 \df x}\nonumber
\,.\eea
The superscript ``naive'' in \eq{C_naive} indicates that this is not yet the final result for the (bare) matching coefficient.

First, we note that we still have to regulate rapidity divergences that arise as $w_2 \to 0$, or equivalently $\myzeta \to 1$.
In our approach, this regulator must only act on the total momentum $k$. The only known regulator in the literature that fulfills this constraint
is the exponential regulator~\cite{Li:2016axz}, which amounts to inserting a factor $\exp[2 \tau e^{-\gamma_E} k^0]$ into the integral.
In our parameterization, this regulating factor reads
\begin{align} \label{eq:exp_regulator}
 \exp(2 \tau e^{-\gamma_E} k^0) &
 = \exp[-\tau e^{-\gamma_E} ( w_1 p_2^+ + w_2 p_1^-) ]
 \to \exp\Bigl[- \frac{\tau e^{-\gamma_E}\qpsq}{\omega_b (1-\myzeta)(1-x)} \Bigr]
\,,\end{align}
where in the last step we neglect the $w_1$ term that is not required to regulate the $w_2\to0$ limit and use the momentum fraction $\omega_b$ of eq.~\eqref{eq:omegadef}.
Since \eq{exp_regulator} vanishes exponentially as $\myzeta\to1$ and $x\to 1$, it regulates all rapidity divergences in the $p_2$-collinear sector.
We identify the rapidity regularisation scale as
\beq
\nu=\frac{1}{\tau}\,,
\eeq
as $\tau$ has inverse mass dimensions.

Secondly, the TMDFF is defined as the purely collinear limit of the cross section, but  the above matrix element still contains overlap with the soft factor.
Its subtraction is referred to as zero-bin subtraction~\cite{Manohar:2006nz}.
In the case of the exponential regulator, this is equivalent to dividing by the bare soft function.
The soft function was calculated at N$^3$LO in \refcite{Li:2016ctv} and confirmed by us in ref.~\cite{Ebert:2020yqt} from which we take its bare expression.

With the above manipulations, we obtain the actual bare matching coefficient as
\begin{align} \label{eq:C}
\tilde{\cK}_{jj^\prime}\Bigl(\myzeta,\bt,\eps, \frac{\nu}{\omega_b}\Bigr) &
 = \int\! \frac{\df^{d-2}\qt \, e^{-\img \qt \cdot \bt}} {\Omega_{d-3}(\qpsq)^{d/2-2}/2}  \int_0^1 \df x \, \df w_2 \, \delta\bigl[q_T^2 - Q^2 (1-\myzeta) w_2 (1-x)  \bigr]
 \\\nn&\quad\times
 \lim_{\tau\to0} \frac{1}{\myzeta^2} \frac{ \exp\Bigl[- \frac{\tau}{\omega_b} e^{-\gamma_E}\qpsq \frac{1}{(1-\myzeta)(1-x)} \Bigr]}{S(b_T, \eps, \tau)}
 \bnlim  \frac{\df\hat\eta_{\bar j+h\rightarrow j^\prime+X}}{\df w_1 \df w_2 \df x} \bigg|_{w_1= -\frac{1-\myzeta}{\myzeta}}
\,,\end{align}
where we already take the limit $\tau\to0$ which must be taken before $\eps\to0$.

The last step is to relate the above partonic coefficient function to its counterpart in production kinematics.
As outlined in \sec{crossing}, the two are related by
\begin{align}
 \frac{\df\hat\eta_{i+h\rightarrow j+X}}{\df w_1 \df w_2 \df x}
 = \myzeta^{3-2\eps} \frac{\cN_i}{\cN_{ij}^{\rm production}}
  \frac{\df\hat\eta_{i+h\rightarrow j+X}}{\df w_1 \df w_2 \df x} \bigg|_\mathrm{analyt. cont.}
\,.\end{align}
To perform the analytic continuation in the above equation the necessary information on the original partonic coefficient function must however be retained as explained in \sec{crossing}.

Since all other ingredients in \eq{C} agree with the corresponding calculation of the matching kernel $\cI_{ij}$ of the TMDPDF in \refcite{Ebert:2020lxs,Ebert:2020yqt}, the relation between the two can be written compactly as 
\begin{align} \label{eq:C_to_I}
 \tilde{\cK}_{jj'}\Bigl(\myzeta,\vec b_T, \eps, \frac{\nu}{\omega_b}\Bigr)
 = \myzeta^{1-2\eps} \frac{\cN_i}{\cN_{ij}^{\rm production}}
  \tilde\cI_{jj'}\Bigl(\frac{1}{\myzeta}, \vec b_T, \eps, \frac{\nu}{\omega_b}\Bigr) \bigg|_\mathrm{analyt. cont.}
\,.\end{align}
It only remains to absorb all leftover UV and IR singularities into suitable counterterms, which we perform in the $\overline{\text{MS}}$ scheme.
This yields the renormalized matching kernel as
\begin{equation} 
\label{eq:I_ren_qT}
 \tilde{\cK}_{ij}\Bigl(\myzeta,\vec b_T,\mu,\frac{\nu}{\omega_b}\Bigr)
 = \sum_{j'}  \int_\myzeta^1 \frac{dz^\prime}{z^\prime}\Gamma_{jj'}(z^\prime)  \tilde Z_B^i(\mu,\tau,Q) \hat Z_{\as}(\mu)
    \tilde{\cK}_{ij'}^{}\left(\frac{z^\prime}{\myzeta},\vec b_T,\epsilon,\frac{\nu}{\omega_b}\right)
\,,\end{equation}
where the factor $ \hat Z_{\as}(\mu,\eps)$ implements the UV renormalization of the strong coupling constant,
$\Gamma_{jj'}$ absorbs all IR poles and corresponds to the redefinition of the bare fragmentation function $d_{H/j}$ in terms of its renormalized counterpart,  and the TMDFF counterterm $\tilde Z_B^i$ absorbs all leftover UV divergences.
$\Gamma_{jj'}$ can be obtained from the time-like splitting functions,
while $\tilde Z_B^i$ can be predicted from the renormalization group equation governing the TMDFF.
These steps are identical to the ones for the TMDPDF, and all required details can be found in appendix A of \refcite{Ebert:2020lxs},
up to replacing the spacelike splitting functions $P_{ij}$ by their timelike counterparts $P_{ij}^T$~\cite{Chen:2020uvt,Mitov:2006ic,Moch:2007tx,Almasy:2011eq}.

The perturbative matching kernel for the manifestly scheme-independent TMDFF of eq.~\eqref{eq:B_to_f} is simply obtained by
\beq
\label{eq:kernelnonu}
\tilde{\cK}_{jj^\prime }^{\text{TMD}}\Bigl(\myzeta,\vec b_T,\mu,\omega_b^2 \Bigr) =  \tilde{\cK}_{jj^\prime }\Bigl(\myzeta,\vec b_T,\mu,\frac{\nu}{\omega_b} \Bigr)  \sqrt{S(b_T, \mu, \nu)},
\,.\eeq
where $\zeta = \omega_b^2$ is the Collins-Soper scale.

\subsection{Results}
\label{sec:results}

We expand the renormalized matching kernels perturbatively as
\begin{align}\label{eq:logdef}
 \tilde{\cK}_{jj^\prime }\Bigl(z,\vec b_T,\mu,\frac{\nu}{\omega_b} \Bigr) &
 = \sum_{\ell=0}^\infty\left(\frac{\as}{\pi}\right)^\ell \sum_{n=0}^{2\ell} \sum_{m=0}^{\ell} \tilde{\cK}^{(\ell,m,n)}_{jj'}(z) L_b^n L_\omega^m
\,.\end{align}
The logarithms in \eq{logdef} are defined as
\begin{align}
 L_b = \ln\frac{b_T^2 \mu^2}{4 e^{-2\gamma_E}}
 \,,\qquad L_\omega = \ln\frac{\nu}{\omega_b}
 \,.\end{align}
The logarithmic structure of \eq{logdef} is entirely governed by the renormalization group equations of the TMDFF,
which we have verified as an important check of our results.
The key new result of this article is the nonlogarithmic boundary term in \eq{logdef},
\begin{align}
\tilde{\cK}^{(\ell)}_{jj'}(z) \equiv \tilde{\cK}^{(\ell,0,0)}_{jj'}(z)
\,.\end{align}
These coefficients have already been calculated at NNLO in \refscite{Echevarria:2015usa, Echevarria:2016scs, Luo:2019bmw, Luo:2019hmp}, with which we find perfect agreement,  while our result at N$^3$LO is new.
As in the case of the TMDPDF, we find that it can be entirely expressed in terms of harmonic poly logarithms (HPLs)~\cite{Remiddi:1999ew} up to weight five.
We provide the full result for \eq{logdef}, and the corresponding result including the soft factor as defined in \eq{kernelnonu}, in the ancillary files of this submission.
We also provide the expansion of the kernels both in $z$ as well as in $\bar{z} = 1-z$ up to 40 orders in the expansion.
These expansions can be patched together to obtain a fast and precise numerical evaluation of the kernels.

We have performed several checks on ours results.
First, in \eq{I_ren_qT}, we have used counterterms predicted from known anomalous dimensions, rather than simply absorbing all appearing divergences in counterterms. As consequence, divergences up to $1/\eps^6$ had to cancel in the process.
In this manner, we also confirm the result of \refcite{Chen:2020uvt} for the timelike splitting functions at three loops,
which noted a discrepancy for $P_{qg}^{T,(2)}$ compared to the previous results of \refcite{Almasy:2011eq},
but otherwise agree with previous determinations of the timelike splitting functions~\cite{Mitov:2006ic,Moch:2007tx,Almasy:2011eq}.

We have also checked that the TMDFF obeys the same eikonal limit as the TMDPDF~\cite{Echevarria:2016scs, Lustermans:2016nvk, Billis:2019vxg},
\begin{align} \label{eq:eikonal_limit}
 \lim_{z\to1} \tilde \cK_{ij}^{(3)}(z) &
 = \frac{\gamma_{2}^r}{64} \, \delta_{ij} \, \cL_0(1-z)
\,,\end{align}
where $\gamma_2^r$ is the three-loop coefficient of the rapidity anomalous dimension in the appropriate color representation $r$.
Explicit expressions for it can be found in Eq.~(9) in \refcite{Li:2016ctv}.

Concerning the partonic coefficient function $\hat\eta_{ij}$, we had already verified in \refscite{Ebert:2020unb,Ebert:2020yqt} that the inclusive integral over all final state kinematics for the soft limit of the coefficient function yields the first term in the threshold expansion of the corresponding inclusive cross section~\cite{Anastasiou:2014lda,Anastasiou:2014vaa,Anastasiou:2015ema,Mistlberger:2018etf,Duhr:2020seh}.
Furthermore, in refs.~\cite{Dulat:2017prg,Dulat:2018bfe} a threshold expansion of the differential perturbative coefficient function for Higgs boson production was performed.
We checked that the first four terms in the threshold expansion of the collinear limit of the limit of $\hat\eta_{ij}$ used here matches the collinear expansion of the threshold expansion of refs.~\cite{Dulat:2017prg,Dulat:2018bfe}.

\begin{figure*}
 \centering
 \includegraphics[width=0.49\textwidth]{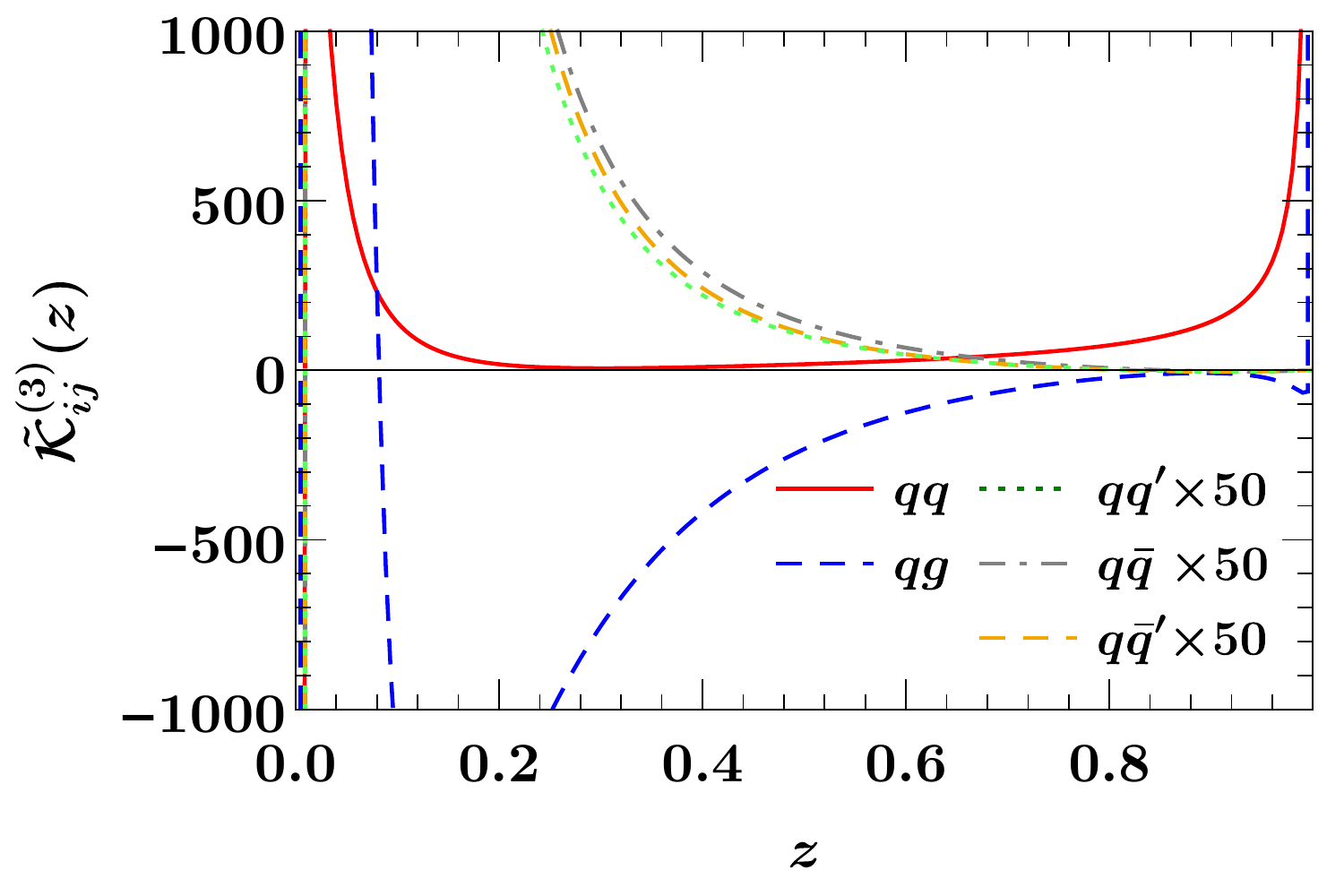}
 \hfill
 \includegraphics[width=0.49\textwidth]{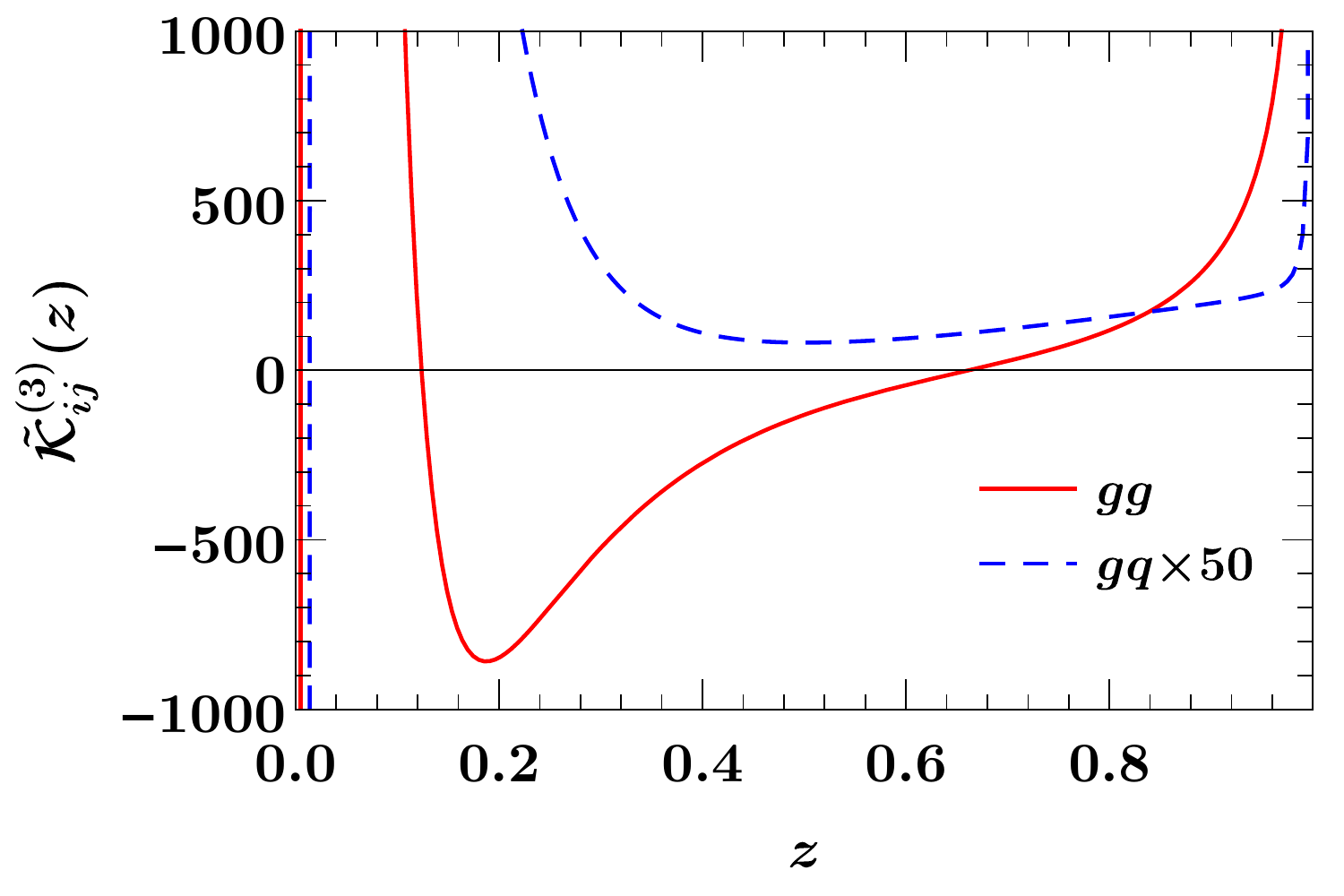}
 \caption{The N$^3$LO TMD fragmentation function boundary term $\tilde{ \cK}^{(3)}_{ij}(z)$ as a function of $z$. The matching kernels in all channels entering the quark fragmentation (left) and the gluon fragmentation (right) are displayed. For illustration purposes the kernels have been rescaled as indicated.}
 \label{fig:C3}
\end{figure*}

In \fig{C3} we illustrate our results by showing the three-loop matching kernel $\cK^{(3)}_{ij}(z)$ in all quark channels (left) and gluon channels (right).
The different channels have been rescaled as indicated in the figure to account for their different magnitudes.

For completeness, in \app{smallx} we present the $\myzeta\to0$ limit of the kernels, both for the quark and the gluon TMD fragmentation functions.
These results are interesting for the study of the high energy behavior of TMDFFs, similar to studies of the small-$x$ behavior of TMDPDFs in \refscite{Balitsky:2015qba,Marzani:2015oyb,Balitsky:2016dgz,Xiao:2017yya,Luo:2019bmw,Ebert:2020yqt}.
Note that the timelike TMDFF shows a double-logarithmic series in $\ln\myzeta$, such that the N$^3$LO coefficient contains up to $\as^3 \ln^5 \myzeta$,
in contrast to the single-logarithmic series observed for the spacelike TMDPDF, where one encounters at most $\as^3 \ln^2\myzeta$ at this order.

\FloatBarrier
\section{Conclusions}
\label{sec:conclusion}

We have computed the perturbative matching kernel relating transverse-momentum dependent fragmentation functions (TMDFFs) with longitudinal fragmentation functions at N$^3$LO in QCD, obtaining analytic results for all partonic channels contributing to the quark and unpolarized gluon TMDFF.
These results for this matching kernel, defined in eq.~\eqref{eq:our_dearest_kernel}, are provided as ancillary files together with the arXiv submission of this article.

Our calculation is based on a simple extension of a framework recently developed by us, that allows to expand differential hadronic cross sections efficiently in the collinear limit~\cite{Ebert:2020lxs}.
This method was developed in detail in \refcite{Ebert:2020lxs} for the collinear expansion of differential hadron collider production cross sections.
We have demonstrated explicitly how they are related to DIS cross sections via analytic continuation.
By analytically continuing our recent computation of the collinear limit of the gluon fusion Higgs boson and DY production cross section to DIS kinematics, we have obtained the TMDFFs in similar fashion as the $N$-jettiness beam functions and TMDPDFs calculated in \refscite{Ebert:2020lxs,Ebert:2020unb,Ebert:2020yqt}.
Our new results demonstrate once more the potency of this method obtaining universal ingredients arising in the infrared and collinear limits of QCD to an unprecedented level of precision in perturbation theory.

An important check on our calculation lies in the cancellation of all infrared and ultraviolet poles against suitable counterterms.
Since these counterterms can be fully predicted using known anomalous dimension, this provides a highly nontrivial check.
In particular, it involves the cancellation of infrared divergences against the QCD mass factorisation counterterm comprised of time-like splitting functions.
Thus, as a by product, our calculation confirms the recent results for the NNLO timelike splitting function \refcite{Mitov:2006ic,Moch:2007tx,Almasy:2011eq,Chen:2020uvt}, in particular the correct result in the $qg$ channel first obtained in \refcite{Chen:2020uvt}.

There are several phenomenological applications of our results.
Firstly, the TMDFFs obtained in this paper constitute the last missing ingredient to describe the singular structure of the transverse momentum distribution of QCD radiation in color-singlet decays at N$^3$LO.
They also enable the resummation of transverse momentum distributions at N$^3$LL$^\prime$ accuracy, both in $e^+e^-$ annihilation and Higgs decay to quarks or gluons as well as in SIDIS.
In particular they allow for the calculation of the jet functions for the Energy-Energy Correlator (EEC) and the Transverse EEC jet functions in the back-to-back limit~\cite{Moult:2018jzp,Gao:2019ojf} at N$^3$LO. For the case of the EEC this allows to push the resummation accuracy to N$^3$LL$^\prime$ which constitutes the most accurate resummation carried for an event shape to date. We carry out this calculation in \refcite{Ebert:2020sfi}.

Our method to expand cross sections around the collinear limit in the final state can be used to calculate higher order terms in the collinear expansion. 
Such higher order terms would allow one to study the structure of factorization beyond leading power for IRC safe observables in $e^+e^-$ annihilation and Higgs decay~\cite{Beneke:2002ph,Pirjol:2002km,Beneke:2002ni,Bauer:2003mga,Freedman:2013vya,Freedman:2014uta,Moult:2016fqy,Moult:2017jsg,Goerke:2017lei,Balitsky:2017gis,Feige:2017zci,Moult:2017rpl,
Chang:2017atu,Beneke:2018rbh,Moult:2018jjd,Moult:2019mog} as well as the appearance of subleading power rapidity divergences~\cite{Moult:2017xpp,Ebert:2018gsn,Moult:2019vou}.
Furthermore, they would provide data to validate the resummation of power suppressed logarithms~\cite{Moult:2018jjd,Moult:2019uhz}.
It would also be interesting to explore the application of the methods developed here and in \refcite{Ebert:2020lxs} to TMDFFs involving a jet measurement~\cite{Kang:2020xyq,Makris:2020ltr,Arratia:2020ssx}.
\\
\\
{\bf Note:} While this article was under completion, an independent calculation was made available on the arXiv in~\refcite{Luo:2020epw} based on the method proposed in \refcite{Chen:2020uvt}.
The authors of ref.~\cite{Luo:2020epw} provided an important cross check on intermediary results for genuine two loop contributions in the $\cK_{gq}$ channel  that allowed us to track an error in a routine related to the analytic continuation of the partonic coefficient functions. 
The initial discrepancy was a non-logarithmically enhanced finite and rational term proportional to $(C_A-C_F)\zeta_2\zeta_3$ in the $\cK_{gq}$ and $\cK_{qg}$ channel. 
After this was resolved, we find perfect agreement among all analytic results.

\acknowledgments
We are grateful to the authors of ref.~\cite{Luo:2020epw}, Ming-xing Luo, Tong-Zhi Yang, HuaXing Zhu, Yu Jiao Zhu, for very helpful comparisons.
This work was supported by the Office of High Energy Physics of the U.S. DE-AC02-76SF00515 and by the Office of Nuclear Physics of the U.S.\ DOE under Contract No.\ DE-SC0011090 and within the framework of the TMD Topical Collaboration.
M.E.\ is also supported by the Alexander von Humboldt Foundation through a Feodor Lynen Research Fellowship.

\appendix

\section{SIDIS Factorization at small transverse momentum}
\label{app:factorization}

In this appendix we provide more information on the factorization theorem for SIDIS at small $q_T$.
For concreteness, we focus on the unpolarized photon process
\begin{align}
 P(P_1) \, + \,\gamma(q) \rightarrow H(-P_2) \,+\, X(-k)
\,.\end{align}
The extension to a scattering with a scalar Higgs boson is trivial.
The corresponding matrix element is given by
\begin{align}
 \cM_{P+q \to H+X} = \eps_\mu(q) \braket{H X | J^\mu | P}
\,,\end{align}
where $\eps_\mu(q)$ is the polarization vector of the incoming photon, and $J^\mu$ the QCD current it couples to.
The resulting cross section for this process is given by
\begin{align} \label{eq:xs_factorized_1}
 \df\sigma &
 = \frac{\pi}{4 P_1 \cdot q} \frac{\df^3 P_2}{2 E_2} (-g_{\mu\nu})
   W^{\mu\nu}(q, P_1, P_2)
\,,\end{align}
where following \refcite{Collins:1350496} we have defined the hadronic tensor as
\begin{align} \label{eq:W}
 W^{\mu\nu}(q, P_1, P_2) &= \sum_X \delta^4(P_1 + q + P_2 + k) \braket{P | J^{*\mu} | H X} \braket{H X | J^\nu | P}
\,,\end{align}
and the $-g_{\mu\nu}$ in \eq{xs_factorized_1} arises from averaging over the photon polarizations.
Working in the Breit frame as specified by \eq{framedef}, the hadron momentum $P_2$ can be parameterized as
\begin{align}
 P_2^\mu &
 = \Bigl( \sqrt{ \vec P_{2T}^2 + (x_F Q/2)^2} \,,\,  \vec P_{2T} \,,\, - \frac{Q x_F}{2} \Bigr)
 \nn\\&
 = \Bigl( \frac{Q x_F}{2} \,,\,  \vec P_{2T} \,,\, - \frac{Q x_F}{2} \Bigr) + \cO\biggl(\frac{P_{2T}^2}{Q^2}\biggr)
\,,\end{align}
where $\vec P_{2T}$ is the Euclidean transverse momentum of the outgoing hadron,
and $x_F = - 2 P_2 \cdot q / q^2$ was defined in \eq{xB_xi}.
This immediately yields
\begin{align}
 \frac{\df^3 \vec P_2}{2 E_2} &
 = \frac{\df^2 \vec P_{2T} \df x_F}{2 x_F}  + \cO\biggl(\frac{P_{2T}^2}{Q^2}\biggr)
\,.\end{align}
Suppressing the power corrections and using \eq{xB_xi}, we obtain the differential cross section as
\begin{align} \label{eq:xs_factorized_2}
 \frac{\df\sigma}{\df x_F \df^2 \vec P_{2T}} &= \frac{\pi}{4 q^2} \frac{x_B}{x_F} W^\mu_\mu(q, P_1, P_2)
\,.\end{align}

The factorized hadronic tensor is typically given in the frame where the outgoing hadron has no transverse momentum,
but the photon momentum $q^\mu$ acquires a transverse component $q_T$. At small transverse momentum, the two are related by
(see e.g.~\eq{framerels})
\begin{align}
 \vec P_{2T} = - x_F \qt
\,.\end{align}
The cross section differential in small $\qt$ thus follows from \eq{xs_factorized_2} as
\begin{align} \label{eq:xs_factorized_3}
 \frac{\df\sigma}{\df x_F \df^2\qt} &= \frac{\pi}{4} \frac{x_B x_F}{-q^2} (-W^\mu_\mu)(q, P_1, P_2)
\,.\end{align}
The factorized hadronic tensor is given by~\cite{Collins:1350496}
\begin{align} \label{eq:W2}
 W^{\mu\nu}(q,P_1,P_2) &
 = 8 \pi \aem \, x_F \sum_f (-g_\perp^{\mu\nu}) H_{f\bar f}(q^2, \mu^2)
 \\\nn&\quad\times
   \int\frac{\df^2\bt}{(2\pi)^2} e^{\img \qt \cdot \bt}
   \tilde f_{f}^\TMD(x_B, \bt, \mu, \zeta_a) \tilde D_{H/\bar f}^\TMD(x_F, \bt, \mu, \zeta_b)
\,,\end{align}
where $\aem$ is the electromagnetic coupling constant, and $\zeta_{a,b}$ are the Collins-Soper scales such that $\zeta_a \zeta_b = Q^4$.
Compared to the formulation in \refcite{Collins:1350496}, we have defined the scalar hard function normalized such that
$H_{f\bar f}(q^2,\mu^2) = Q_f^2 [ 1 + \cO(\as)]$, where $Q_f$ is the charge of the quark the photon couples to.
The overall factor of $x_F$ in \eq{W2} compensates for the factor of $1/x_F$ in the definition of the TMDFF, see \eq{def_D}.
Also note that \refcite{Collins:1350496} uses a different variable $z$ for the momentum fraction of the outgoing hadron,
which at small $q_T$ reduces to
\begin{align}
 z_\text{\cite{Collins:1350496}} = - \frac{P_1 \cdot P_2}{P_1 \cdot q}
 = x_F + \cO\biggl(\frac{q_T^2}{Q^2}\biggr)
\,.\end{align}
The TMDPDF $\tilde f$ and TMDFF $\tilde D$ in \eq{W2} are defined with absorbing the soft factor.
For our purpose, it will be more convenient to disentangle the soft from the collinear sectors,
which is easily achieved by using \eq{B_to_f}. Together with \eqs{W2}{xs_factorized_3}, we obtain the desired result
\begin{align} \label{eq:xs_factorized_4}
 \frac{\df\sigma}{\df x_F \df^2\qt} &= (2\pi)^2 \aem \frac{x_B x_F^2}{Q^2} \sum_f H_{f\bar f}(q^2, \mu^2)
 \\\nn&\quad\times
   \int\frac{\df^2\bt}{(2\pi)^2} e^{\img \qt \cdot \bt} \tilde B_{f}(x_B, \bt, \mu, \nu/\omega_a) \tilde D_{H/\bar f}(x_F, \bt, \nu/\omega_b) \tilde S_q(b_T, \mu, \nu)
\,.\end{align}

\section{High-energy limit of the TMD fragmentation function kernels}
\label{app:smallx}
{\normalsize In this appendix, we provide the asymptotic behaviour in the high-energy limit of the boundary term, i.e. the $L_b$ and $L_\omega$ independent term in $\bt$-space, of the TMDFFs kernels.
Here we report only the new results for the small-$z$ limit of the $\cO(\alpha_s^3)$ kernels, normalized by $\left(\frac{\alpha_s}{4\pi}\right)^3$. 
In the high-energy limit, the kernels are enhanced by a double logarithmic series. This is peculiar of the timelike nature of the TMDFF kernels, as their spacelike analog, the TMD beam function kernels, are only single logarithmically enhanced in the small-$z$ limit~\cite{Ebert:2020yqt,Luo:2019bmw}. 
Note that also splitting functions are single logarithmically enhanced in the spacelike case, while they obey a double logarithmic series at small-$z$ in the timelike case~\cite{Moch:2004pa,Vogt:2004mw,Mitov:2006ic,Moch:2007tx,Almasy:2011eq,Chen:2020uvt}.
Therefore, this different behavior in the high-energy limit between spacelike and timelike TMD functions is similar to the small-$x$ behavior of timelike vs spacelike splitting functions.
The high-energy limit $z\to0$ of the kernels $\tilde \cK_{gg}^{(3)}(z)$ and $\tilde \cK_{gq}^{(3)}(z)$ contributing to the gluon TMD fragmentation function is given by\par}
{\footnotesize
\allowdisplaybreaks
\bea
\lim\limits_{z\rightarrow 0}  z \, \tilde \cK_{gg}^{(3)}(z)&=&
	32 C_A^3 \blue{\log^5(z)}+\blue{\log^4(z)}\left[\frac{8008}{27}C_A^3+\frac{176}{27}C_A^2n_f-\frac{896}{27}C_A C_Fn_f\right]
\nn\\&+&
	\blue{\log^3(z)} \left[C_A^3  \left(\frac{57392}{81}-\frac{256}{3}\zeta_2\right)+\frac{2576}{27}C_A^2 n_f+\frac{64}{81}C_An_f^2-\frac{14624}{81}C_A C_F n_f
	-\frac{128}{81}C_F n_f^2\right]
\nn\\&+&
	\blue{\log^2(z)} \left[C_A^3 \left(-\frac{352}{3}\zeta_2+416 \zeta_3+\frac{14792}{27}\right)+C_A^2 n_f \left(\frac{128}{3}\zeta_2+\frac{11408}{81}\right)+\frac{368}{81}C_A n_f^2
\right. \nn\\ &-& \left.	
    C_A C_F n_f \left(\frac{128}{9} \zeta_2+\frac{18536}{81}\right)+\frac{32}{3}C_F^2 n_f-\frac{736}{81}C_F n_f^2\right]
\nn\\&+&
\blue{\log(z)} \left[ C_A^3 \left(\frac{19984 \zeta_2}{27}+\frac{5632 \zeta_3}{9}+2104 \zeta_4-\frac{344864}{243}\right)-C_A C_F n_f \left(\frac{1504 \zeta_2}{27}+\frac{1280 \zeta_3}{9}-\frac{46280}{243}\right)
\right. \nn\\ &+& \left.
	C_A^2 n_f \left(-\frac{1808 \zeta_2}{27}+\frac{128 \zeta_3}{9}-\frac{23456}{81}\right)+\frac{944}{243}C_A n_f^2
	+C_F^2 n_f\left(\frac{512 \zeta_3}{9}-\frac{176}{3}\right)+\frac{4640}{243}C_F n_f^2\right]
\nn\\&+&
   C_A^3 \left(448 \zeta_3 \zeta_2-\frac{34640}{81}\zeta_2-\frac{14576}{9}\zeta_3+3256
   \zeta_4+4336 \zeta_5-\frac{1348136}{243}\right)
\nn\\&-&
   C_A n_f^2 \left(\frac{32 \zeta_3}{9}+\frac{13064}{729}\right)+C_F n_f^2 \left(\frac{64
   \zeta_3}{9}+\frac{26128}{729}\right) +C_F^2
   n_f \left(\frac{1376 \zeta_3}{9}+\frac{416 \zeta_4}{3}-\frac{842}{3}\right)
\nn\\&+& 
	C_AC_F n_f \left(\frac{1928 \zeta_2}{27}-\frac{776 \zeta_3}{3}-\frac{4360\zeta_4}{9}+\frac{424732}{729}\right)
\nn\\&+&
	C_A^2 n_f \left(-\frac{1568}{81}\zeta_2+\frac{1576}{9}\zeta_3+120 \zeta_4-\frac{115420}{729}\right)
\,,\\[3mm]
\lim\limits_{z\rightarrow 0} z \, \cK_{gq}^{(3)}(z)&=&
	-\frac{368}{27}C_A^2 \blue{\log^4(z)} + \blue{\log^3(z)}\left[-\frac{7328}{81} C_A^2+\frac{224}{81}C_A n_f+\frac{320}{81}C_F n_f\right]
	\nn\\&+& 
\blue{\log^2(z)} \left[-C_A^2 \left(\frac{416}{9}\zeta_2+\frac{580}{9}\right)+ C_AC_F\left(\frac{128}{3}\zeta_2-\frac{32}{3}\right)-\frac{832}{81}C_A n_f+\frac{1280}{81}C_F n_f\right]
\nn\\&+& 
	\blue{\log(z)}\left[C_A^2 \left(-\frac{2320}{27}\zeta_2+\frac{128}{3}\zeta_3+\frac{7180}{243}\right)+ C_A C_F
	\left(\frac{256}{9}\zeta_2-64 \zeta_3+\frac{500}{9}\right)
\right. \nn\\ &+& \left.
	C_A n_f \left(\frac{284}{81}-\frac{128}{27}\zeta_2\right)+C_F n_f \left(\frac{56}{3}-\frac{128 \zeta_2}{27}\right)\right] -C_A  n_f \left(\frac{128}{81}\zeta_2+\frac{16}{9}\zeta_3-\frac{64}{27}\right)
   	\nn\\&+& C_A C_F \left(\frac{1520 \zeta_2}{27}+\frac{16 \zeta_3}{9}-48 \zeta_4+\frac{257}{27}\right)+C_F n_f \left(-\frac{512}{81}\zeta_2+\frac{32 \zeta_3}{9}+\frac{2576}{729}\right)
	\nn\\&+& 
   	C_A^2 \left(-\frac{1184 \zeta_2}{27}+\frac{488}{9}\zeta_3-\frac{1216}{9}\zeta_4+\frac{76196}{243}\right)
\,.\eea
\par}%
{For the quark channels, the high-energy limit $z\to0$ of the kernels $\tilde \cK_{qi}^{(3)}(z)$ contributing to the quark TMD fragmentation function is given by\par}
{\footnotesize
\allowdisplaybreaks
\bea
\lim\limits_{z\rightarrow 0}  z \, \tilde \cK_{qq}^{(3)}(z)&=&\lim\limits_{z\rightarrow 0}  z \,\tilde \cK_{q\bar{q}}^{(3)}(z)=\lim\limits_{z\rightarrow 0}  z \,\tilde \cK_{qq^\prime}^{(3)}(z)=\lim\limits_{z\rightarrow 0}  z \,\tilde \cK_{q\bar{q}^\prime}^{(3)}(z) 
\nn\\&=&
	-\frac{368}{27} C_A C_F \blue{\log^4(z)}+\blue{\log^3(z)} \left[\frac{128 C_F n_f}{27}-\frac{7168
	C_A C_F}{81}\right]
\nn\\&+&
	\blue{\log^2(z)} \left[C_A C_F \left(-\frac{32 \zeta_2}{9}-\frac{7316}{81}\right)+\frac{64 C_F^2}{3}+\frac{32
	C_F n_f}{9}\right]
\nn\\&+&
	\blue{\log(z)} \left[C_A C_F \left(-\frac{1808}{27}\zeta_2+\frac{128}{3}\zeta_3+\frac{5656}{243}\right)+C_F^2 \left(\frac{64}{9}\zeta_2-64 \zeta_3+\frac{644}{9}\right)
\right. \nn\\ &+& \left.
	C_F n_f \left(\frac{1280}{81}-\frac{64}{9}\zeta_2\right)\right]+C_A C_F \left(-\frac{8800 }{81}\zeta_2+56 \zeta_3-\frac{736}{9}\zeta_4+\frac{256078}{729}\right)
\nn\\&+&
	C_F^2 \left(\frac{3136}{27}\zeta_2+\frac{16}{9}\zeta_3-\frac{304}{3}\zeta_4-\frac{631}{27}\right)+C_F n_f \left(-\frac{16}{9}\zeta_2-\frac{1424}{729}\right)
\,,\\[3mm]
\lim\limits_{z\rightarrow 0} z \,\tilde \cK_{qg}^{(3)}(z)&=& 32 C_A^2 C_F \blue{\log^5(z)}+\blue{\log^4(z)} \left[\frac{7816 C_A^2 C_F}{27}-\frac{16 C_A C_F
   n_f}{27}-\frac{512 C_F^2 n_f}{27}\right]
\nn\\&+&
   \blue{\log^3(z)} \left[C_A^2 C_F \left(\frac{7376}{9}-\frac{2560 \zeta_2}{9}\right)+C_A C_F^2
   \left(\frac{1792 \zeta_2}{9}-\frac{1360}{9}\right)+ \frac{1856}{81}C_A C_F n_f
\right. \nn\\ &-& \left.
  \frac{2464}{27} C_F^2 n_f\right]+ \blue{\log^2(z)} \left[C_F^3
   \left(-64 \zeta_2+\frac{128 \zeta_3}{3}+\frac{208}{3}\right)+C_A^2 C_F \left(\frac{1184}{3}\zeta_3-\frac{1984}{3}\zeta_2+\frac{8608}{9}\right)
\right. \nn\\ &+& \left.
   C_A C_F^2 \left(\frac{1888 \zeta_2}{3}-\frac{64
   \zeta_3}{3}-\frac{1532}{3}\right)+C_A C_F n_f \left(\frac{448 \zeta_2}{9}+\frac{76}{81}\right)-C_F^2 n_f \left(\frac{128 \zeta_2}{3}+\frac{3592}{27}\right)\right]
\nn\\&+&
   \blue{\log(z)} \left[C_A^2 C_F \left(\frac{3776 \zeta_2}{3}+\frac{3136 \zeta_3}{3}+1408 \zeta_4-\frac{123892}{81}\right)-C_F^2 n_f \left(\frac{640 \zeta_2}{9}+\frac{224 \zeta_3}{3}-\frac{1030}{27}\right)
\right. \nn\\ &+& \left. 
   C_AC_F^2 \left(-\frac{1360 \zeta_2}{3}-\frac{80 \zeta_3}{3}+\frac{488 \zeta_4}{3}+\frac{701}{9}\right)-C_AC_F n_f \left(\frac{2192 \zeta_2}{27}+\frac{160 \zeta_3}{3}+\frac{33692}{243}\right)
   \right. \nn\\ &-& \left. 
   C_F^3 \left(\frac{224 \zeta_2}{3}+336 \zeta_3-\frac{1600 \zeta_4}{3}+\frac{173}{3}\right)\right] + C_A C_F n_f \left(\frac{5432 \zeta_2}{81}+\frac{352 \zeta_3}{3}+\frac{344 \zeta_4}{9}-\frac{25300}{729}\right)
\nn\\&+&
	C_A^2 C_F \left(\frac{608}{3} \zeta_3\zeta_2+\frac{21944}{27}\zeta_2-\frac{7588}{9}\zeta_3+\frac{5870}{3}\zeta_4+\frac{7456}{3}\zeta_5-\frac{3650707}{729}\right)
\nn\\&+&
    C_A C_F^2 \left(992 \zeta_3 \zeta_2-\frac{48568 \zeta_2}{27}-1764
	\zeta_3+\frac{830}{3}\zeta_4+\frac{6800}{3}\zeta_5+\frac{10141}{9}\right)
\nn\\&+&
	C_F^3 \left(-\frac{2240}{3}\zeta_3 \zeta_2+608 \zeta_2+\frac{2888}{3}\zeta_3+796 \zeta_4-416 \zeta_5-\frac{4715}{3}\right)
\nn\\&-&
	C_F^2 n_f \left(\frac{2992}{27}\zeta_2+\frac{832}{9}\zeta_3+\frac{112}{3}\zeta_4-\frac{37885}{729}\right)
\,.\eea
\par}%

The expressions for the high energy limit ${z\rightarrow 0}$ up to $\cO(z^{40})$, as well as that for the threshold limit $z\to 1$ up to $\cO((1-z)^{40})$, can be found for all channels in electronic form in the ancillary files of this work.

\addcontentsline{toc}{section}{References}
\bibliographystyle{jhep}
\bibliography{../refs}

\end{document}